\documentclass[superscriptaddress,twocolumn]{revtex4-2}

\usepackage[T1]{fontenc}
\usepackage[utf8]{inputenc}

\usepackage[yyyymmdd]{datetime} %

\usepackage[dvipsnames]{xcolor}
\usepackage{hyperref}
\hypersetup{
  breaklinks=true,
  colorlinks=true,
  allcolors=BlueViolet,
}

\usepackage{graphicx}
\graphicspath{{./figures/}} %
\usepackage{subfigure}

\usepackage[export]{adjustbox} %
\usepackage{makecell}
\usepackage{float}
\usepackage{braket,bm,amssymb}
\newcommand{\op}[2]{\ket{#1}\!\!\bra{#2}}
\usepackage{booktabs}
\usepackage{mathtools}

\DeclarePairedDelimiter{\abs}{\lvert}{\rvert}

\usepackage{stmaryrd}
\newcommand{\params}[1]{\llbracket #1 \rrbracket}

\usepackage{amsmath}
\usepackage[capitalize]{cleveref}
\makeatletter
\AddToHook{cmd/appendix/before}{%
    \crefalias{section}{appendix}%
    \crefalias{subsection}{appendix}
}
\makeatother
\usepackage{amsthm,thmtools}
\newtheorem{theorem}{Theorem}

\newtheorem{condition}{Condition}

\crefname{condition}{Condition}{Conditions}

\usepackage[inline]{enumitem} %
\setlist[enumerate]{label=(\arabic*)} %
\setlist{nosep} %

\usepackage{tikz}
\usepackage{xcolor}
\usepackage[compat=0.6]{yquant}
\usetikzlibrary{quantikz2}

\newcommand{\wn}{\setwiretype{n}}
\newcommand{\basis}[1]{\rstick{\!\scriptsize{$#1$}}}

\definecolor{red}{rgb}{0.702, 0, 0.008}

\newcommand{\GTAR}{Global Technology Applied Research, JPMorganChase, New York, NY 10017, USA}
\newcommand{\QUK}{Quantinuum, Partnership House, Carlisle Place, London SW1P 1BX, UK}

\begin{document}
\interfootnotelinepenalty=10000 %
\count\footins = 1000 %

\title{Injection is all you need: parallel universal logic in high-rate concatenated self-dual codes}
\author{Joseph Sullivan}
\affiliation{\GTAR}
\author{David Amaro}
\affiliation{\QUK}
\author{Michael A. Perlin}
\affiliation{\GTAR}
\date{\today}

\begin{abstract}
  We propose a quantum architecture that achieves addressable and parallel universal fault-tolerant logical in high-rate codes.
  The codes are obtained from the flexible concatenation of small-distance high-rate self-dual CSS codes.
  Addressable and parallel universal logic on the resulting $K$ logical qubits is injected by fault-tolerantly preparing and then consuming diverse logical $K$-qubit CSS stabilizer states and logical $\ket{H}^{\otimes K}$ global magic states.
  Flag-at-origin is used to prepare the stabilizer resource states while a concatenated cultivation protocol prepares the global magic states.
  These resource states are consumed by Steane-QEC or Knill-QEC teleportation gadgets, which simultaneously perform quantum error correction.
  Any quantum algorithm with $T$-depth $D_T$ can be executed by consuming $12D_T$ or fewer resource states.
  We benchmark the fault-tolerant preparation of the resource states for several two-level concatenations of the $\params{k+4,k,2}$ H codes and the Hamming $\params{15,7,3}$ code.
  In particular, we find that the concatenated cultivation offers at least one order of magnitude spacetime reduction compared to cultivation in the surface code.
\end{abstract}

\maketitle

\tableofcontents

\section{Introduction}

High-rate quantum error-correcting (QEC) codes are enticing for their asymptotically low physical overheads as a quantum memory.
However, it is often unclear how to use these codes for quantum \emph{computation}.
Strategies include generalized lattice surgery with suitably-constructed ancilla code patches~\cite{xu2024constantoverhead, xu2025fast, papadopoulos2026stabilizer, gu2026qgpu}, logical gates based on code automorphisms~\cite{sayginel2025fault, koh2026entangling, malcolm2026computing, yang2026spacetime}, and the direct circuit-level construction of logical gates that address a chosen subset of the logical qubits in a code block~\cite{goto2024highperformance, yoshida2025concatenate, malcolm2025computing, tansuwannont2026construction}.
A fundamental problem for high-rate codes is that they generally suffer from an addressability problem~\cite{guyot2025addressability, krishna2025tradeoffs, chakraborty2026nogo}, which seemingly limits computational bandwidth (active circuit width) by limiting the number of logical qubits that can be processed simultaneously, forcing logically disjoint gates to be serialized.

\begin{figure*}
  \centering
  \includegraphics{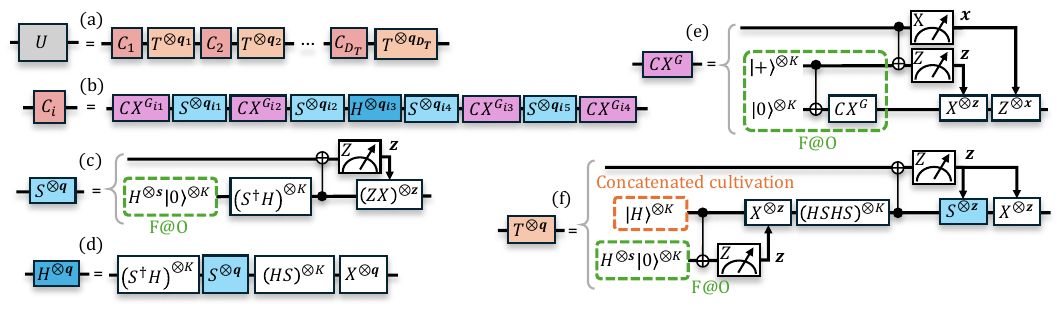}
  \caption{
  The injection architecture employs the concatenation of small-distance self-dual SWEL codes~\cite{perlin2026quantum, tansuwannont2026construction} that encode $k>1$ logical qubits. These codes have a transversal physical implementation of $S^{\otimes k}$ and $H^{\otimes k}$ logical gates.
  (a) Compilation of any unitary with $T$-depth $D_T$ into $D_T$ layers of Clifford gates $C_i$ and products $T^{\otimes \bm{q}_i}$ of $T$ gates acting on subsets of the $K$ logical qubits of the final concatenated block.
  At most $12D_T$ resource states are required to inject all logical gates and simultaneously perform quantum error correction (QEC): $9$ for each Clifford layer and $3$ for each $T$ layer.
  (b) Bruhat's decomposition of any Clifford circuit into 9 Clifford gates of three types: $CX^{G}$ is a product of CNOT gates that act sequentially on an ordered list of qubit pairs $G$, $S^{\otimes \bm{q}}$ is a product of logical $S$ gates, and $H^{\otimes \bm{q}}$ is a product logical $H$ gates.
  (c) Steane-QEC injection of $S^{\otimes \bm{q}}$ by consuming a tensor product $H^{\otimes \bm{q}}\ket{0}^{\otimes K}$ resource state prepared with flag-at-origin (F@O) on one concatenated code block.
  (d) Steane-QEC injection of $H^{\otimes \bm{q}}$.
  (e) Knill-QEC injection of $CX^{G}$ by consuming a two-block resource state prepared with F@O.
  (f) Steane-QEC injection of $T^{\otimes \bm{q}}$ by consuming a one-block $\ket{H}^{\otimes K}$ resource state prepared with concatenated cultivation and a one-block $H^{\otimes \bm{q}}\ket{0}^{\otimes K}$ resource state.
  }
  \label{fig:main}
\end{figure*}

In this work, we take a different route that avoids the addressability problem: rather than building addressable logical gates into the code, we inject all but a handful of logical gates via Steane-QEC and Knill-QEC teleportation gadgets that consume fault-tolerantly prepared resource states.
A byproduct of these injections is that we simultaneously perform QEC and mitigate leakage noise by teleporting the logical information into fresh qubits.
Our scheme is designed for a specific and practically-motivated family of codes: concatenated codes built from self-dual CSS seed codes, called self-dual with equivalent logicals (SWEL)~\cite{perlin2026quantum, tansuwannont2026construction}, that have transversal global logical Hadamard ($H$) and phase ($S = \mathop{\mathrm{diag}}(1,i)$) operators.
Specifically, an $\params{n,k}$ SWEL code is a self-dual CSS code for which $\bar{H}^{\otimes k} = H^{\otimes n}$ and $\bar{S}^{\otimes k} \cong_Z S^{\otimes n}$, where $\bar H$ and $\bar S$ denote logical $H$ and $S$ operators, and $\cong_Z$ denotes equality modulo single-qubit $Z$ gates~\footnote{All self-dual CSS codes with odd block length $n$ are SWEL, as are self-dual CSS codes with even block length, but at least one logical Pauli-$X$ operator with support on an odd number of physical qubits~\cite{tansuwannont2026construction}.}.
This restriction enables the fault-tolerant construction of resource states suitable for highly parallel universal quantum computation with high-rate concatenated codes~\cite{yoshida2025concatenate}.
We sketch out the general scheme in \cref{fig:main}.

Our approach to executing Clifford circuits is inspired by Ref.~\cite{zheng2020constant}, which showed that an arbitrary logical Clifford circuit can be injected by consuming a constant number of resource states, one per stage of a Bruhat decomposition \cite{maslov2018shorter}.
In that work the requisite resource states are generically non-CSS and are produced by a costly distillation protocol.
Our key observation is that the availability of global $H$ and $S$ gates allows every stage of the decomposition to be reduced to a CSS resource state; an arbitrary $k$-qubit logical Clifford can then be injected using at most 9 CSS resource states.
Fault-tolerant preparation circuits for these CSS states are found without distillation using the flag-at-origin method of Ref.~\cite{forlivesi2025flag}.
We emphasize that the tools underlying our construction of Clifford gates depend only on the use of SWEL codes (or, more generally, fault-tolerant logical $\bar{H}^{\otimes k}$ and $\bar{S}^{\otimes k}$ gates).
In particular, neither code concatenation (as the method for code construction) nor flag-at-origin (as the method for CSS resource state preparation) are required for our fault-tolerant Clifford gates.
Nonetheless, code concatenation is essential for our construction of a high-rate fault-tolerant magic resource state factory.

To complete a universal gate set we must inject a non-Clifford gate, for which we use consumable logical $\ket{H}^{\otimes k}$ magic states.
Here the transversal $\bar{H}^{\otimes k}$ gate plays a second, distinct role: because $\bar{H}^{\otimes k}$ is transversal, it can be measured fault-tolerantly, providing a built-in flagging mechanism that detects errors in the magic-state preparation.
On its own, a single such parity measurement suppresses magic-state infidelity to only $O(p^2)$, since the measurement of $\bar{H}^{\otimes k}$ can only resolve the parity of faulty logical qubits.
Combining this flagging mechanism with code concatenation, however, lets us cross-verify independent parity measurements at successive levels and drive the probability of an undetectable logical error down to $O(p^{2^\ell})$ at concatenation level $\ell$.
We refer to this procedure as concatenated cultivation.
It is closely related to the protocols explored in Ref.~\cite{goto2014stepbystep, dasu2025breaking, Jones_2013, Knill_2005}.

\section{Logical Clifford gates}
\label{sec:Clifford}

Quantum teleportation is known to be a universal logical primitive~\cite{gottesman1999demonstrating}.
Consider a $k$-qubit unitary $U = \sum_{\bm a,\bm b\in\set{0,1}^k} U_{\bm a\bm b} \op{\bm a}{\bm b}$ that we wish to apply to the $k$ logical qubits of a single code block, initially in the logical state $\ket\psi$.
This unitary can be injected by the circuit
\begin{align}
  \includegraphics[valign=c]{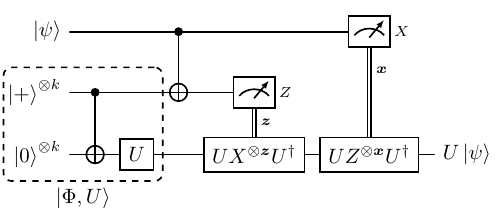},
  \label{eq:gate_telep}
\end{align}
where we identify a sub-circuit that prepares the state $\ket{\Phi,U} = \frac1{2^k} \sum_{a,b\in\set{0,1}^k} U_{ab} \ket{ba}$, the bitstrings $\bm x,\bm z\in\set{0,1}^k$ are logical measurement outcomes, and throughout this work we adopt the notation $V^{\otimes\bm v} = \bigotimes_{j=1}^{\abs{\bm v}} V^{v_j}$ for any single-qubit unitary $V$ and integer vector $\bm v = (v_1, v_2, \cdots)$.
If $U$ is Clifford, then $U X^{\otimes\bm z} U^\dag$ and $U Z^{\otimes\bm x} U^\dag$ are Pauli strings.

At face value, \cref{eq:gate_telep} merely converts the problem of applying $U$ to $\ket\psi$ into the problem of applying $U$ to the second register of $k$ Bell pairs.
However, if we can prepare $\ket{\Phi,U}$ directly, then \cref{eq:gate_telep} tells us how to consume this state to inject $U$ to an arbitrary $k$-qubit state $\ket\psi$.
We have thus reduced the problem of applying $U$ to that of preparing the resource state $\ket{\Phi,U}$.

While fault-tolerantly preparing $\ket{\Phi,U}$ may be difficult in general, if $U$ is Clifford then we can simplify our task via the Bruhat decomposition~\cite{maslov2018shorter} $U = \prod_{j=1}^9 U_j$ into Cliffords $(U_1, U_2, \cdots, U_9)$ with type signature $(\texttt{CX},\texttt{S},\texttt{CX},\texttt{S},\texttt{H},\texttt{S},\texttt{CX},\texttt{S},\texttt{CX})$ as in \cref{fig:main}(b).
Here a Clifford of type $\texttt{CX}$, $\texttt{H}$, or $\texttt{S}$ can be exclusively built, respectively, out of CNOT, $H$, or $S$ gates.
Ref.~\cite{zheng2020constant} thereby proposed to inject an arbitrary Clifford $U_j$ by consuming the resource states $\ket{\Phi,U_j}$, for which the authors designed suitable fault-tolerant distillation protocols.
The idea is to use faulty state preparation and then concatenate with classical codes $[n_z,k_z,d_z]$ and $[n_x, k_x, d_x]$.
If $d_x, d_z > d$, then $n_xn_z$ noisy copies of the CSS state $\ket{\psi}$ can be distilled into $k_xk_z$ copies of $\ket{\Phi,U_j}$ with fault distance $d$.
The strength of this approach is clear when many copies of the same state $\ket{\Phi,U_j}$ are required. 
However, the Bruhat decomposition does not generically require many copies of each resource state.
To produce a single copy of $\ket{\Phi,U_j}$ via distillation, the optimal choice is to use the classical repetition code $[d+1,1,d+1]$, which means that each distillation factory requires $n(d+1)^2$ qubits.

Here, we keep the idea of consuming resource states to inject Cliffords, but replace distillation by the distillation-free flag-at-origin method of Ref.~\cite{forlivesi2025flag}.
An immediate obstruction to this approach is that flag-at-origin can only fault-tolerantly prepare CSS states that are stabilized by Pauli-$X$ and Pauli-$Z$ strings, whereas the resource states $\ket{\Phi,U_j}$ are generally non-CSS.
However, we can overcome this obstruction if we restrict our attention to SWEL codes, which have fault-tolerant logical global $S^{\otimes k}$ and $H^{\otimes k}$ gates~\cite{perlin2026quantum, tansuwannont2025clifford} that act on all logical qubits.
\begin{theorem}
  Given a quantum CSS code that encodes $k$ logical qubits, the following primitives suffice to implement any $k$-qubit Clifford unitary on the logical qubits of one code block:
  \begin{enumerate}[label=(\alph*)]
    \item a logical $S^{\otimes k}$ gate within one code block,
    \item a $H^{\otimes k}$ gate within one code block,
    \item the preparation of arbitrary logical CSS states.
  \end{enumerate}
  Here a CSS state $\ket{\mathcal{G}}$ is a state that can be uniquely defined by a set of stabilizers, $g\ket{\mathcal{G}} = \ket{\mathcal{G}} \forall g\in\mathcal{G}$, that consist of only Pauli-$X$ and Pauli-$Z$ strings.
  \label{thm:css_clifford}
\end{theorem}
\begin{proof}
  Decompose the $k$-qubit Clifford unitary $U = \prod_{j=1}^9 U_j$ into a sequence of nine circuits with type signature $(\texttt{CX},\texttt{S},\texttt{CX},\texttt{S},\texttt{H},\texttt{S},\texttt{CX},\texttt{S},\texttt{CX})$, using the Bruhat decomposition \cite{maslov2018shorter}.
  If each of the circuits $U_j$ is fault-tolerant (and sufficient QEC is performed in between, to form suitable extended rectangles~\cite{gottesman2010introduction}), then the unitary $U$ is fault-tolerant.
  There are three types of circuits to consider in this decomposition.
  \begin{enumerate}
    \item CNOT (\texttt{CX}) type: Each \texttt{CX}-type circuit $V$ consists of only CNOT gates, so the corresponding resource state $\ket{\Phi,C}$ can be obtained by applying logical CNOT gates to logical Bell states of a CSS code, which are altogether CSS states.

    \item Phase (\texttt{S}) type: Without loss of generality, each \texttt{S}-type circuit can be written in the form $S^{\otimes\bm s} Z^{\otimes\bm z}$ for some pair of bitstrings $\bm s,\bm z\in\set{0,1}^k$.
      The logical $Z^{\otimes\bm z}$ gate can be implemented directly with physical single-qubit Pauli-$Z$ gates, which is fault-tolerant.
      For the logical $S^{\otimes\bm s}$ gate, by assumption we can fault-tolerantly prepare tensor products of logical $\ket{0}$ and $\ket{+}$ states, and in particular the state $H^{\otimes\bm s} \ket{0}^{\otimes k}$, which is $\ket{+}$ on the qubits where $\bm s$ is 1, and $\ket{0}$ elsewhere.
      In turn, we can use global operations to convert $H^{\otimes\bm s} \ket{0}^{\otimes k}$ into the state
      \begin{align}
        S^{\otimes\bm s} \ket{+}
        \simeq (H S^\dag)^{\otimes k}
        H^{\otimes\bm s} \ket{0}^{\otimes k}
      \end{align}
      where $\simeq$ denotes equality up to global phase.
      This state is $\ket{i} = \ket{0} + i\ket{1}$ on the qubits where $\bm s$ is 1, and $\ket{+}$ elsewhere.
      Transversally injecting the rotation-gate teleportation gadget
      \begin{align}
        \includegraphics[valign=c]{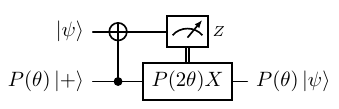}
        \label{eq:phase_telep}
      \end{align}
      where $P(\theta) = \exp(-i\frac{\theta}{2} Z)$ and in particular $P(\frac{\pi}{2})=S$, fault-tolerantly applies the gate $S^{\otimes\bm s}$ to one code block~\footnote{Note that Ref.~\cite{zheng2020constant} provides a method to implement \texttt{S}-type circuits using (addressable) logical CNOT gates (i.e., \texttt{CX}-type circuits) and a global $S$ gate, which altogether requires only CSS resource states to implement.
      However, the gate teleportation method provided here requires fewer ancilla code blocks.}.

    \item Hadamard (\texttt{H}) type: Without loss of generality, every \texttt{H}-type circuit can be written in the form $H^{\otimes \bm h}$ for some bitstring $\bm h\in\set{0,1}^k$.
      By decomposing $H = X S_Y$, where $S_Y = \exp(-i\frac{\pi}{4}Y)$ is an $S$ gate in the $Y$ basis, and in turn $S_Y \simeq (S H) S (H S^\dag)$, we can expand
      \begin{align}
        H^{\otimes\bm h}
        = X^{\otimes\bm h} S_Y^{\otimes\bm h}
        \simeq X^{\otimes\bm h}
        (SH)^{\otimes k}
        S^{\otimes\bm h}
        (HS^\dag)^{\otimes k}.
      \end{align}
      The addressable $H^{\otimes\bm k}$ gate thus reduces to physical Pauli-$X$ gates, global $S^{\otimes k}$ and $H^{\otimes k}$ gates, and an addressable $S^{\otimes\bm h}$ gate, all of which are fault-tolerant.
  \end{enumerate}
\end{proof}
The takeaway from \cref{thm:css_clifford} is that by using Knill-QEC gadgets for CNOT layers (\cref{eq:gate_telep} and \cref{fig:main}(e)) and Steane-QEC gadgets for single qubit Cliffords (\cref{eq:phase_telep} and \cref{fig:main}(c,d)) we can perform arbitrary $k$-qubit Clifford gates by consuming at most 9 CSS resource states.
Furthermore, the process of consuming resource states passively performs quantum error correction and leakage mitigation in tandem with the execution of the circuit.

As a final point, we comment on the implementation of Clifford gates between code blocks, which by the Bruhat decomposition reduces to the task of implementing a CNOT circuit on multiple code blocks.
A CNOT circuit that addresses a pair of $\params{n,k,d}$ SWEL code blocks can be injected by treating them as a single $\params{2n,2k,d}$ code block and applying the Knill-QEC injection of \cref{fig:main}(e).
An alternative route for identical $\params{n,k,d}$ SWEL blocks is to combine their transversal inter-block CNOT with (injected) intra-block CNOTs, as in Refs.~\cite[Figure 9(a)]{goto2024highperformance} and \cite[Figure 7(a)]{liu2025coniq}.
Finally, inter-block CNOTs may be implemented via the construction of homomorphic CNOTs~\cite{huang2023homomorphic} or generalized lattice surgery~\cite{xu2024constantoverhead, xu2025fast, papadopoulos2026stabilizer, gu2026qgpu}.
Unlike the CNOT circuits that are injected with a CSS resource state, however, inter-block CNOTs applied with techniques that rely on addressability typically need to be divided into layers of disjoint logical gates, and implemented one layer at a time, hindering the parallelism of the injection architecture. 
It is therefore important to minimize such inter-block CNOT gates, for example by developing a suitable compiler or using an architecture in which the primary purpose of these gates is to teleport information into and out of a large SWEL processing block.

\section{Logical non-Clifford gates}
\label{sec:non-clifford}

To complete a computationally universal gate set, we need to show how to perform a logical non-Clifford gate.
Recall that any single-qubit rotation $U$ can be approximated to arbitrary precision by a sequence of alternating $T$ and $H$ gates: $U \approx U_L (T^\pm H T^\pm H T^\pm \cdots H T^\pm) U_R$, where $U_L$ and $U_R$ are Clifford.
Below, we will show how $\ket{H}^{\otimes k}$ can be used as a resource state for addressable $T$ gates within a code block, where $\ket{H}$ is a $+1$ eigenstate of $H$.
We will then discuss how to prepare high-fidelity $\ket{H}^{\otimes k}$ states in concatenated SWEL codes.
In particular, we show that repeated use of logical Hadamard measurements gives a state preparation protocol for $\ket{H}^{\otimes k}$ where the probability of an undetectable logical error is $O(p^{2^\ell})$ in the physical error rate $p$ and concatenation level of $\ell$.

\subsection{Universality using $\ket{H}^{\otimes k}$}
\label{sec:ty_gate}

\begin{theorem}
  The primitives in \cref{thm:css_clifford} together with ancillary code blocks in logical $\ket{H}^{\otimes k}$ states suffice to inject an arbitrary Clifford+$T$ circuit on the $k$ logical qubits of one code block.
  \label{thm:t_gates}
\end{theorem}
\begin{proof}
  For the purposes of this proof is sufficient for us to show how to implement an addressable logical gate of the form $T^{\otimes\bm t}$ with $\bm t = (t_j: j\in\mathbb{Z}_k) \in \set{0, 1}^k$.

  As a warm up, setting $\theta=\frac{\pi}{4}$ in \cref{eq:phase_telep} recovers a gadget that consumes a $\ket{T} = SHS\ket{H}$ state to inject a $T$ gate.
  A $T^{\otimes k}$ gate can therefore be applied to the $k$ qubits of a code block by consuming a $\ket{T}^{\otimes k} = S^{\otimes k} H^{\otimes k} S^{\otimes k}\ket{H}^{\otimes k}$ state in parallel using a transversal CNOT gate and an addressable logical $SX$ correction.

  A logical $T^{\otimes\bm t}$ gate can similarly be applied by using the gadget in \cref{eq:phase_telep} with the resource state $T^{\otimes\bm t} \ket{+}^{\otimes k}$.
  To this end, we expand
  \begin{align}
    T^{\otimes\bm t} \ket{+}^{\otimes k}
    = (SHSH)^{\otimes k} \ket{H,\bm t},
  \end{align}
  where
  \begin{align}
    \ket{H,\bm t} \equiv \bigotimes_j \ket{H,t_j},
    &&
    \text{where}
    &&
    \ket{H, \bm t} =
    \begin{cases}
      \ket{H} & t = 1 \\ \ket{0} & t = 0
    \end{cases}.
  \end{align}
  In turn, we can prepare $\ket{H,\bm t}$ from $\ket{H}^{\otimes k}$ by consuming one $H^{\otimes\bm{t}}\ket{0}^{\otimes k}$ state:
  \begin{align}
    \begin{quantikz}
      \lstick{$\ket{H}^{\otimes k}$} & \ctrl{1} & \gate{X}                        & \rstick{$\ket{H,\bm t}$} \\
      \lstick{$H^{\otimes\bm{t}}\ket{0}^{\otimes k}$}       & \targ{}  & \meter{}\basis{Z}\wire[u][1]{c} & \wn
    \end{quantikz}.
    \label{eq:TYI}
  \end{align}
  In total, $T^{\otimes\bm t} \ket{+}^{\otimes k}$ state can be prepared from one $\ket{H}^{\otimes k}$ state and one $H^{\otimes\bm t}\ket{0}^{\otimes k}$ state.
  Consuming the $T^{\otimes\bm t} \ket{+}^{\otimes k}$ state using the gadget in \cref{eq:phase_telep}, which requires a transversal CNOT gate and an addressable logical $SX$ correction, then applies a logical $T^{\otimes\bm t}$ gate as depicted in \cref{fig:main}(f).
\end{proof}

We remark that the sign of individual $T$ gates in $T^{\otimes k}$ can be flipped by sandwiching a $T$ gate with transversal $X$ gates.
Control over local signs $\bm\sigma\in\set{+1,-1}^k$ in $T^{\otimes\bm\sigma}$ also enables an injection of $T^{\otimes\bm o}$ for any odd integers $\bm o\in(2\mathbb{Z}+1)^k$ by decomposing $T^{\otimes\bm o} \simeq T^{\otimes\bm o^T}\cdot Z^{\otimes\bm o^Z}$, where $\simeq$ denotes equality up to global phase, and the exponents $\bm o^T\in\set{+1,-1}^k$ and $\bm o^Z\in\set{0,1}$ are defined by
\begin{align}
  o^T_j &=
  \begin{cases}
    +1 & o_j = 1 \mod 4 \\
    -1 & o_j = 3 \mod 4
  \end{cases},
  \\
  o^Z_j &=
  \begin{cases}
    0 & o_j = \pm 1 \mod 8 \\
    1 & o_j = \pm 3 \mod 8
  \end{cases}.
\end{align}

\subsection{High fidelity $\ket{H}^{\otimes K}$ states via concatenated cultivation}
\label{sec: H prep}

Being able to prepare $\ket{H}^{\otimes K}$ fault-tolerantly on the final $\params{N,K,D}$ $L$-level concatenated code is of paramount importance.
Here, we present a method for producing these states in concatenated codes $\mathbf{C} = C_1 \circ C_2 \dots \circ C_L$, where each $C_j$ is a SWEL code with parameters $\params{n_j,k_j,d_j}$.
The idea, which we refer to as \textit{concatenated cultivation} is to iteratively use the zero-level distillation (ZLD) procedure described below to create $k$ magic states with an infidelity that is $O(p^{2^\ell})$ in the physical error rate $p$ and concatenation level of $\ell$.
Conveniently, these codes also satisfy the conditions of \cref{thm:css_clifford}, making them suitable for universal fault-tolerant quantum computation.

Similar approaches to creating high fidelity magic states in concatenated codes were studied in \cite{dasu2025breaking, goto2014stepbystep}.
Without specifying these details, the $O(p^{2^\ell})$ suppression quoted above is really a statement about the probability of an undetectable logical error occurring.
We elaborate on this point in \cref{sec: decoding comments} and \cref{sec:examples}, where we study concatenated cultivation circuits for specific example codes.

\subsubsection{Zero-level distillation (ZLD)}

Suppose we prepare the state $\ket{H}^{\otimes k}$ in a SWEL code with parameters $\params{n,k,d}$.
Since $H^{\otimes k}$ is transversal and a stabilizer of $\ket{H}^{\otimes k}$, we can measure it fault-tolerantly and post-select on measurement outcomes of $M_{H^{\otimes k}} = +1$, thereby boosting the fidelity of $\ket{H}^{\otimes k}$.
Schematically, after appending a round of code-stabilizer measurements the net state preparation circuit looks like
\begin{align}
  \includegraphics[valign=c]{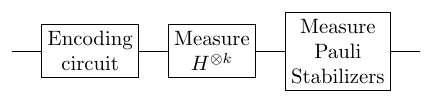}
  \label{eq:H-FT}
\end{align}
where the measurement of $H^{\otimes k}$ can be performed by transversally applying the circuit
\begin{align}
  \begin{quantikz}[row sep=.5em, column sep=1em]
    \lstick{$\ket{+}$} & \ctrl{1} & \meter{}\basis{X} & \wn \\
                       & \gate{H} &                   &
  \end{quantikz}
  =
  \begin{quantikz}[row sep=.5em, column sep=.5em]
    \lstick{$\ket{+}$} &                    & \ctrl{1} & \meter{}\basis{X} & \wn \\
                       & \gate{T_Y^\dagger} & \phase{} & \gate{T_Y}        &
  \end{quantikz}.
  \label{eq:MH}
\end{align}
This trick has been used previously in topological and concatenated codes~\cite{Chamberland_2020, gidney2024magicstatecultivationgrowing, goto2014stepbystep, dasu2025breaking}.
In the case of $k=1$, Ref.~\cite{Chamberland_2020} explains how $O(d)$ Hadamard measurements can be used to obtain a logical infidelity of $O(p^{\lfloor( d+1\rfloor/2)})$.
The situation for $k>1$ is different because a parity measurement of $H^{\otimes k}$ is insensitive to, for example, an even number of $Y$ errors in $\ket{H}^{\otimes k}$.
Repeated measurements of $H^{\otimes k}$ can therefore at best suppress the infidelity of $\ket{H}^{\otimes k}$ to $O(p^2)$.
For this reason, we only require that in the circuit to initially prepare $\ket{H}^{\otimes k}$, all individual faults are detected by the $H^{\otimes k}$ or stabilizer measurements.

To this end, we identify the potentially ``problematic'' Pauli errors that would pass the checks in \cref{eq:H-FT} after the initial preparation of $\ket{H}^{\otimes k}$.
Any Pauli error with destabilizer factors is caught by the code stabilizer measurements, so we need only consider pure logical errors.
As an illustrative example, consider the pure logical error $X_1 Z_2 Y_3$, which acts on $\ket{H}^{\otimes k}$ as
\begin{align}
  \big(X_1 Z_2 Y_3\big) \ket{H}^{\otimes k}
  = \ket{+\tilde{H}}\ket{-\tilde{H}}\ket{-H}\ket{H}^{\otimes k-3}
  \label{eq: Error}
\end{align}
where $\ket{-H}$ is the $-1$ eigenstate of $H$, and $\ket{\pm\tilde{H}} \propto \ket{H} \pm \ket{-H}$ is the $\pm1$ eigenstate of $\tilde{H} = (X-Z)/\sqrt{2}$.
A $+1$ measurement of $H^{\otimes k}$ fails to project onto $\ket{H}^{\otimes k}$ if any two of $X_1$, $Z_2$, or $Y_3$ are present, leading to a general criterion for detectability in \cref{eq:H-FT}:
\begin{condition}[Detectability Condition]\label{cond:detection}
  In order for a mid-circuit Pauli error to be detectable by the stabilizer measurements in \cref{eq:H-FT}, it must be equivalent either to a single (pure logical) Pauli-$X$, a single Pauli-$Z$, or a product of an odd number of Pauli-$Y$ operators.
\end{condition}
If a non-fault-tolerant $\ket{H}^{\otimes k}$ state preparation circuit possess error mechanisms that violate \cref{cond:detection}, then flag gadgets can be added to catch these errors.
We discuss the addition of flag gadgets in more detail in \cref{sec:state prep appendix}.
We provide an encoding circuit that satisfies \cref{cond:detection} for the $\params{15,7,3}$ Hamming code in \cref{fig:hamming encoder} of \cref{sec: Benchmarking magic states}, and encoding circuits for an entire family of $\params{k+4,k,2}$ H codes in \cref{sec: H codes}.

\subsubsection{Concatenated cultivation}

In concatenated codes we can overcome the $O(p^2)$ barrier by applying ZLD at each level of concatenation to verify parity for independent batches of logical states.
Cross-referencing the measurement outcomes from the different levels of concatenation allows us to bound the logical error rate.
In order for cross-referencing to work in a concatenated code $\mathbf{C} = C_1 \circ C_2 \dots \circ C_L$, each code $C_j$ in the sequence must have transversal global $H^{\otimes k_j}$ gate.
This method generalizes that in Ref.~\cite{dasu2025breaking} to the concatenation of any SWEL codes equipped with circuits that satisfy \cref{cond:detection}.

It will be helpful to define some terminology and notation: we refer to the logical qubits at $\ell$ levels of concatenation as level-$\ell$ qubits.
That is, level-$0$ qubits are physical qubits, level-$1$ qubits are the logical qubits of $C_1$, level-$2$ qubits are the logical qubits of $C_1 \circ C_2$, and so on.
We will use integer subscripts on a quantum state, as in $\ket{\psi}_\ell$, to denote that it is expressed as a state of level-$\ell$ qubits.
In fact, we will only need to consider $\ell\le2$ below, as the code $C_1\circ C_2\circ C_3$ can be obtained by concatenating $(C_1\circ C_2)$ with $C_3$, so the generalization of our discussions below to the case of $\ell>2$ follows by induction.

\begin{figure}
  \centering
  \includegraphics[width=\linewidth]{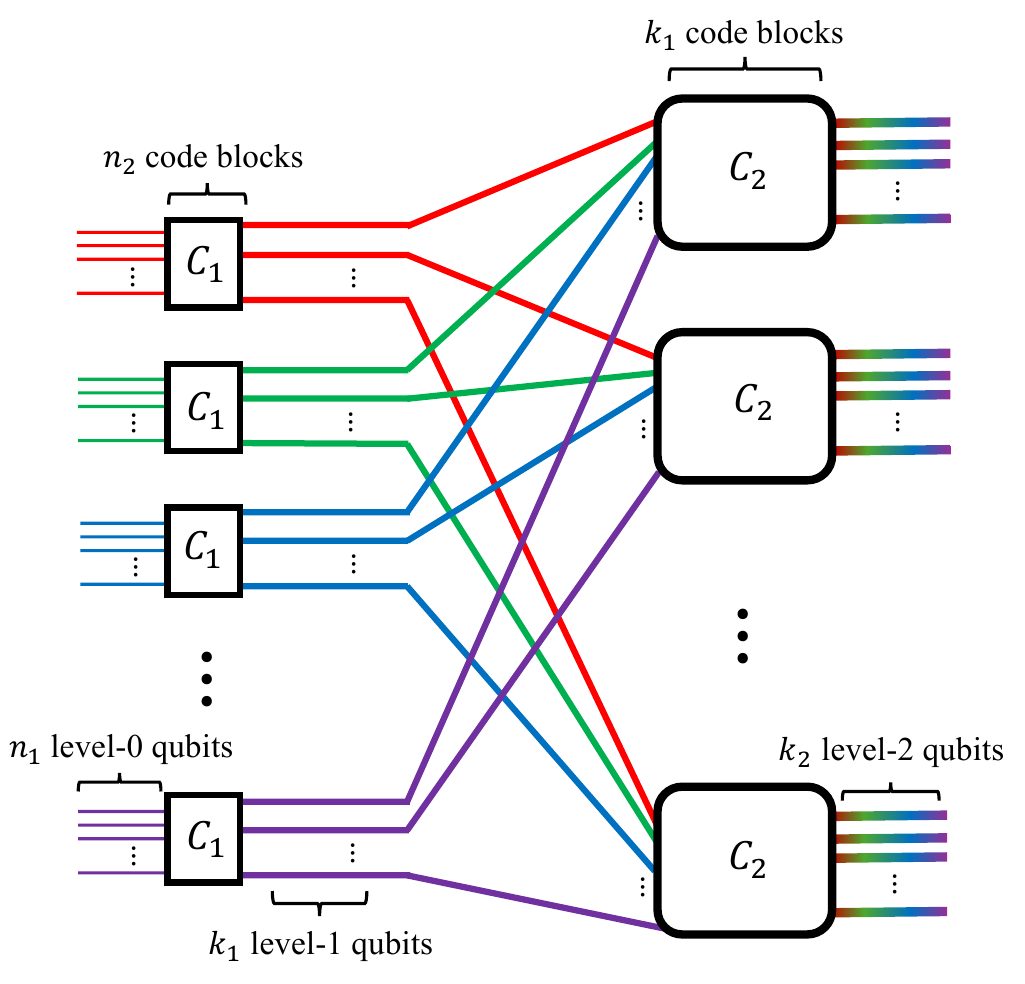}
  \caption{Generalized tree style concatenation yielding $k_1k_2$ level-2 qubits: $n_2$ blocks of $C_1$ provide the level-1 qubits which are encoded into $k_1$ blocks of $C_2$. In a given $C_1$ block each of the $k_1$ level-1 qubits is allocated to a distinct $C_2$ block.}
  \label{fig:concat graph}
\end{figure}

Consider the codes $C_1$ and $C_2$, with parameters $\params{n_1,k_2,d_1}$ and $\params{n_2,k_2,d_2}$, concatenated into the code $\mathbf{C} = C_1 \circ C_2$ with parameters $\params{N,K,D} = \params{n_1n_2,k_1k_2,d_1d_2}$.
Specifically, we consider a generalization of tree-style concatenation, visualized in \cref{fig:concat graph}: each logical (level-1) qubit of a $C_1$ code block is assigned to a distinct $C_2$ code block, such that in total we concatenate $n_2$ copies of a $C_1$ code blocks with $k_1$ copies of a $C_2$ code block.

To prepare $\ket{H}_2^{\otimes K}$, we first prepare $k_2$ copies of $\ket{H}_1^{\otimes k_1}$ using ZLD with $C_1$, and prepare $n_2-k_2$ copies of a $C_1$ code block initialized into $\ket{0}_1^{\otimes k_1}$
We thereby prepare, in effect, $k_1$ copies of the state $\ket{H}_1^{\otimes k_2} \ket{0}_1^{\otimes(n_2-k_2)}$.
We then consider an encoding circuit for $C_2$, which nominally maps the input state $\ket{H}_\ell^{\otimes k_2} \ket{0}_\ell^{\otimes(n_2-k_2)}$ to the encoded state $\ket{H}_{\ell+1}^{\otimes k_2}$ using only level-$\ell$ Hadamard gates, CNOT gates, and the readout or preparation of level-$\ell$ ancilla qubits in the $X$ and $Z$ bases.
The $k_1$-fold tensor product of this encoding circuit similarly maps $k_1$ copies of $\ket{H}_1^{\otimes k_2} \ket{0}_1^{\otimes(n_2-k_2)}$ onto $k_1$ copies of $\ket{H}_2^{\otimes k_2}$ using level-1 gates that are transversal in $C_1$.
The encoding circuit for $C_2$ is required to satisfy \cref{cond:detection} for compatibility with ZLD.

So far, we have encoded level-1 qubits into level-2 qubits, noisily preparing $k_1$ copies of $\ket{H}_2^{\otimes k_2}$.
We now need to measure Hadamard operators $H^{\otimes k_2}$ for ZLD at the second level of concatenation.
To this end, we nominally consider using one level-1 ancilla qubit to measure $H^{\otimes k_2}$ on the logical (level-2) qubits of each $C_2$ code block, with the circuit
\begin{align}
  \includegraphics[valign=c]{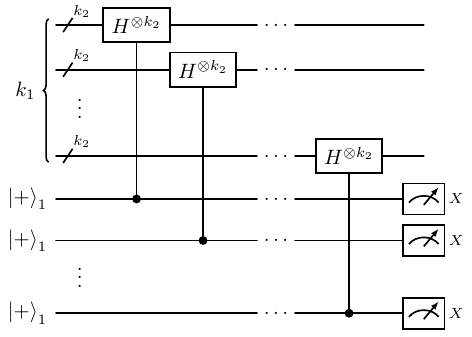}.
  \label{eq:meas_h_start}
\end{align}
This circuit should be thought of as a \emph{formal} object, having a well-defined logical action despite using gates with level-1 control and level-2 target qubits.

Transversality of $H^{\otimes k_2}$ in $C_2$ means that the level-2 $H^{\otimes k_2}$ gate across a $C_2$ code block in \cref{eq:meas_h_start} is equivalent to a $H^{\otimes n_2}$ gate on level-1 qubits.
This push-through makes the $q$-th level-1 ancilla qubit in \cref{eq:meas_h_start} address $q$-th logical qubit of each $C_1$ code block (see \cref{fig:concat graph}), which we can write in terms of transversal controlled-Hadamard gates between the logical (level-1) qubits of $C_1$ code blocks as
\begin{align}
  \includegraphics[valign=c]{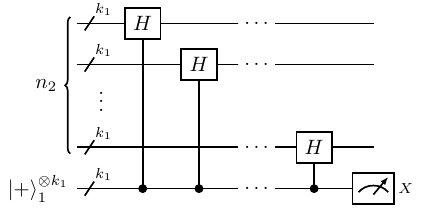}.
  \label{eq:meas_h_start}
\end{align}
Each wire in \cref{eq:meas_h_start} corresponds to one $C_1$ code block, allowing the transversal controlled-Hadamard gates to be implemented by decomposing $H = T_Y Z T_Y^\dag$, as in \cref{eq:MH}.
Finally, to make this circuit have level-1-fault distance 2, we replace the ancilla $C_1$ code block with a pair of $C_1$ code blocks that store $k_1$ Bell pairs,
\begin{align}
  \includegraphics[valign=c]{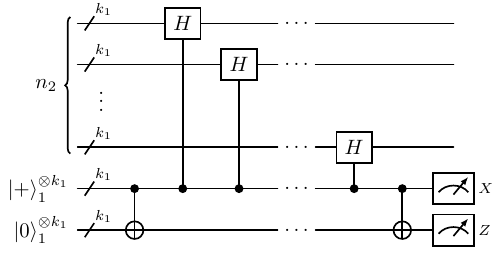}.
  \label{eq:meas_h_end}
\end{align}
Up to half of the transversal two-qubit gates in \cref{eq:meas_h_end} can be applied from the second ancilla code block rather than the first, cutting the circuit depth of \cref{eq:meas_h_end} by nearly half.

To recap, after the first round of ZLD we prepare $k_2$ copies of $\ket{H}^{\otimes k_1}_1$, with a logical error rate $O(p^2)$.
We then run $C_2$ encoding circuits transversally across $C_1$ code blocks to prepare $\ket{H}_2^{\otimes K}$.
Finally, to perform ZLD at the second level of concatenation, which requires measuring a level-2 $H^{\otimes k_2}$ operator in each $C_2$ code block (as in \cref{eq:meas_h_start}), we use the level-1 circuit in \cref{eq:meas_h_end} to measure $H^{\otimes n_2}$ operators distributed across $C_1$ code blocks.
Whereas the parity measurement of $H^{\otimes k_1}$ can be blind to block-local physical errors of weight 2, these errors get caught by the cross-block measurements of $H^{\otimes n_2}$.
Weight-3 errors are similarly caught by the parity measurements for ZLD, altogether giving the post-selected $\ket{H}^{\otimes K}_2$ state a logical error rate of $O(p^4)$.
By induction, iterating this procedure for $L$ levels of concatenation yields magic states with a post-selected logical error rate of $O(p^{2^L})$.
We remark that if the concatenated code is built from seed codes with distance $d_j\ge2$, then the concatenated code distance $D = \prod_{j=1}^L d_j\ge 2^L$, so $O(p^{2^L}) \ge O(p^D)$.

\subsubsection{Code distance versus cultivation distance}
\label{sec: decoding comments}

The $O(p^{2^L})$ probability of a logical error from concatenated cultivation is the probability of a \emph{pure} logical error, with no destabilizer factors.
However, concatenated cultivation still leaves errors of weight $w$ with probability $O(p^w)$.
If the concatenated code distance $D$ is insufficiently large, the true probability of a logical error when \emph{consuming} the cultivated state $\ket{H}_L^{\otimes K}$ in an error-corrected algorithm may instead be limited by the maximal error weight that can be reliably corrected in the code, namely $t = \lfloor(D-1)/2\rfloor$.
The effective logical error rate of $\ket{H}_L^{\otimes K}$ is therefore $O(p^{\min\set{2^L,t+1}})$.
As an example, a weight-3 error in a $\params{90,14,6} = \params{6,2,2}\circ\params{15,7,3}$ code is detectable by stabilizers, but may be decoded incorrectly, increasing the $O(p^4)$ logical error rate of the $\ket{H}_2^{\otimes 14}$ state to $O(p^3)$.

The inability to accommodate the $O(p^{2^L})$ logical error rate is similar to what happens with magic state cultivation on the surface code~\cite{gidney2024magicstatecultivationgrowing}, where the fidelity of the logical state produced in a distance-3 color code is too high to survive decoding in the original code, which must in turn be grown into a larger surface code that can accommodate the low error rate of the state.
The examples benchmarked in this work fall into the same category of codes for which $t+1 < 2^L$.
We provide a detailed discussion of decoding in \cref{sec: noise}, where we explain how to restore, at least partially, the level of error suppression provided by concatenated cultivation by incorporating post-selection into the decoding process.

If $2^L < t+1$, we make full use of concatenated cultivation, but in this case the cultivated state does not exploit the full error-correcting capability of the distance-$D$ code.
The resulting excess distance provides some ``slack'' when designing the cultivation circuit.
Concatenated cultivation iteratively uses a fault-tolerant encoding circuit for each seed code $C_i$ with distance $d_i$, but the circuit fault distance need not equal the distance $d_i$ of that seed code.
Instead, only a fault distance of 2 is needed at each concatenation level, simplifying the task of designing the full concatenated cultivation circuit.
This flexibility can also be used to reduce quantum resources.
Suppose we are targeting a suppression, after decoding, of $O(p^4)$.
The required code distance can be accomplished by a $L = 3$ code with distance-2 seed codes, or a $L = 2$ code with distance-3 seed codes.
The latter code involves fewer concatenations, which generally translates to more favorable post-selection rates and spacetime volumes.
This tradeoff has obvious implications for code choice and architecture design.

\section{Concrete examples}
\label{sec:examples}

In this section, we numerically benchmark the performance of the flag-at-origin and concatenated cultivation state preparation protocols on three example codes with concatenation level $L = 2$: 
\begin{align}
\params{36,4,4} &= \params{6,2,2} \circ \params{6,2,2}, \label{eq:code_A} \\
\params{48,8,4} &= \params{6,2,2} \circ \params{8,4,2}, \label{eq:code_B} \\
\params{96,14,6} &= \params{6,2,2} \circ \params{15,7,3}. \label{eq:code_C}
\end{align}
The seed codes of $\params{36,4,4}$, $\params{48,8,4}$, and $\params{96,14,6}$ are all SWEL codes.
See \cref{sec: Code Definitions} for explicit stabilizers and our choice of logical operators for the seed codes.
The $\params{6,2,2}$ and $\params{8,4,2}$ are the smallest instances of the family of $\params{k+4,k,2}$ H codes (limited to even $k$) \cite{eczoo_quantum_h, Jones_2013}, all of which are all SWEL.
We provide additional benchmarks of cultivated magic states in level-2 concatenated codes built from $\params{k+4,k,2}$ H codes in \cref{sec: H codes}.

\subsection{Noise and decoding}
\label{sec: noise}

Our numerical benchmarks use a depolarizing noise model with the following parameters:
\begin{table}[H]
\centering
\label{tab:noise_model}
\begin{tabular}{@{}|l| l| l| c@{} |}
\hline
Fault           & Channel                       & Applied  & Rate     \\
\hline
Initialisation/Reset      & 1-qubit depolarising          & after         & $p$      \\
Two-qubit gate   & 2-qubit depolarising          & after         & $p$      \\
Measurement    & 1-qubit bit-flip              & before        & $p$      \\
\hline
Single-qubit gate         & 1-qubit depolarising          & after         & $p/100$  \\
Idling (memory)    & 1-qubit depolarising          & during idle   & $p/100$  \\
\hline
\end{tabular}
\end{table}

To compute the logical error rates of a noisy concatenated cultivation circuit, sketched in \cref{fig:high level circuit}, we append a noiseless round of syndrome measurements.
We then decode with a lookup table decoder built using a code-capacity depolarizing noise model for the concatenated code.
The lookup table stores all syndromes generated by errors up to weight $t$.
If the syndrome produced by a sample is not recognized by the decoder, then the decoder declares erasure and the state is rejected, which contributes $O(p^{t+1})$ to the overall post-selection rate.
The scaling is important for overall scalability because decoder-heralded post-selection occurs ``online'' in the sense that the stabilizer measurements that the decoder receives are a by-product of a state injection.
Post-selection here then means aborting an entire circuit, rather than merely restarting a repeat-until-success protocol, so it is crucial that this event is suppressed in $p$.

When $D$ is even, as is the case for all of our example codes, online post-selection has the added benefit of suppressing logical errors from $O(p^{t+1})$ to $O(p^{t+2})$.
In this case, if two errors $E_\mathrm{small}$ and $E_\mathrm{big}$ with weights $\abs{E_\mathrm{small}} \leq t$ and $\abs{E_\mathrm{big}} > t$ have identical syndromes, they can only differ in their logical operator content if $\abs{E_\mathrm{big}} \geq D - t = t + 2$.
The logical error rate of a state prepared by concatenated cultivation therefore becomes $O(p^{\min\set{2^L,t+2}})$.
This can be contrasted with a pure post-selection protocol, wherein no decoder is used and any syndrome violation leads to the state being rejected.
This approach will saturate the $O(p^{2^L})$ suppression but at the expense of an $O(p)$ ``online'' post-selection which limits the depth of executable logical circuits.
As a concrete example, preparing $\ket{H}^{\otimes 4}$ in the $\params{6,2,2} \circ \params{6,2,2}$ code via concatenated cultivation was previously studied in \cite{dasu2025breaking}.
The benchmarking therein was done using pure post-selection on stabilizer measurements with the logical error rate and ``online'' post-selection scaling as $O(p^4)$ and $O(p)$.
For this case, the superior logical error rate cannot be fully taken advantage of because the $O(p)$ post-selection will severely limit the number of magic state injections that can be performed.
When employing a decoder (see \cref{sec: Benchmarking magic states}) these same quantities are more balanced, logical error rate and ``online'' post-selection scale as $O(p^3)$ and $O(p^2)$, allowing for deeper logical circuits.

\subsection{Benchmarking magic states}
\label{sec: Benchmarking magic states}

In this section we describe the encoding circuit for $\ket{H}^{\otimes k}$ in more detail, and present numerical performance results for Clifford proxy circuits which mimic the structure of the aforementioned circuits.

\subsubsection{$\ket{H}^{\otimes k}$ encoding circuits}

\begin{figure}
  \centering
  \includegraphics[width=\linewidth]{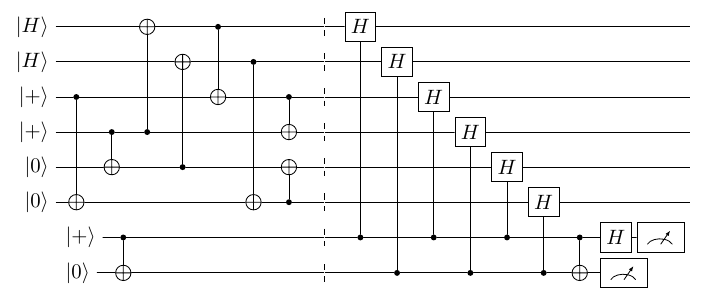}
  \caption{
    FT $\ket{H}^{\otimes 2}$ state prep circuit for $\params{6,2,2}$.
    The dashed line separates the encoding circuit from the Hadamard measurement gadget.
  }
  \label{fig:C6 encoder}
\end{figure}

In order to prepare $\ket{H}^{\otimes K}_2$ in our concatenated codes, we first need to identify encoding circuits for $\ket{H}^{\otimes k}_1$ in each of the seed codes.
A strategy for arbitrary seed codes is to start with a general unitary encoding circuit, and then add flag gadgets to catch all single faults that violate \cref{cond:detection}.
We found that the family of $\params{k+4, k,2}$ H codes admit encoding circuits that satisfy \cref{cond:detection} without flag gadgets (see \cref{sec: H codes}).
\cref{fig:C6 encoder} shows the corresponding encoding circuit (followed by a Hadamard measurement gadget) for the $\params{6,2,2}$ H code.

\begin{figure*}[t]
  \centering
  \includegraphics[width=.9\linewidth]{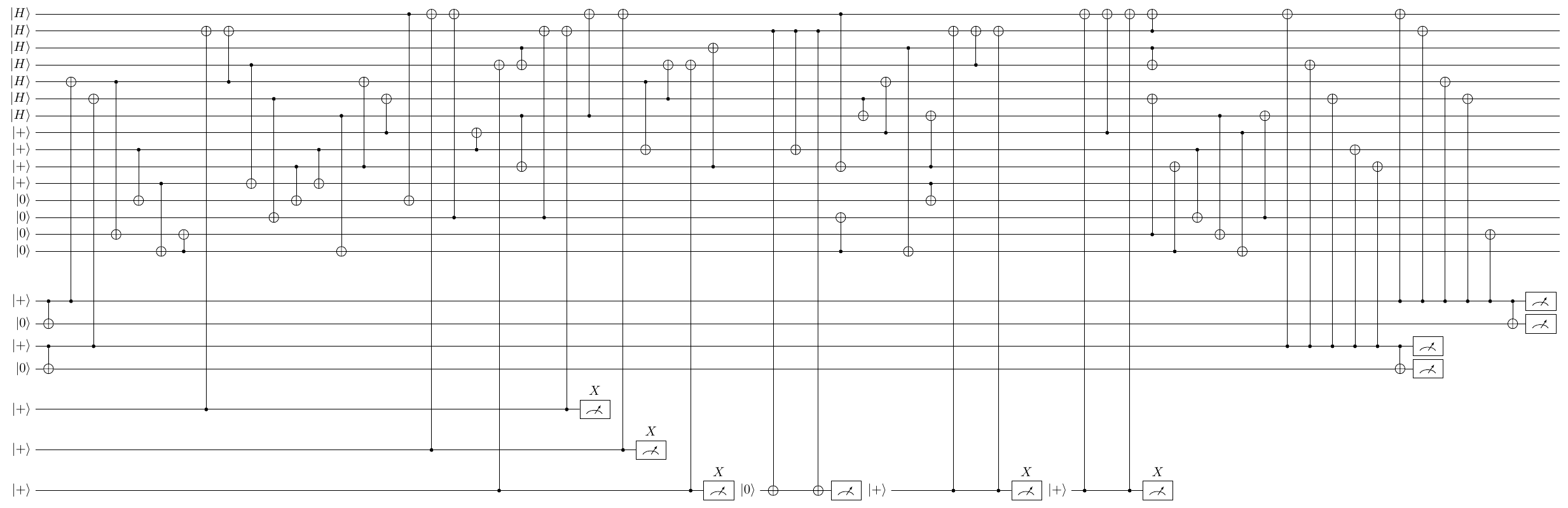}
  \caption{
    Flagged encoder for $\ket{H}^{\otimes 7}$ in the $\params{15,7,3}$ code obeying \cref{cond:detection}.
    Appending Hadamard and stabilizer measurement gadgets makes the circuit FT.
    Measured-out qubits are post-selected on.
    See \cref{sec:state prep appendix} for details about the construction of this circuit.
  }
  \label{fig:hamming encoder}
\end{figure*}

For the $\params{15,7,3}$ Hamming code, we were unable to find an unflagged encoding circuit that satisfies \cref{cond:detection}.
We therefore start with a compact general encoding circuit, produced using the the Munich Quantum Toolkit (MQT)~\cite{mqt}, and use flag gadgets to catch ``problematic'' errors that violate \cref{cond:detection}, such as that in \cref{eq: Error}.
We thereby arrive at the encoding circuit in \cref{fig:hamming encoder}.
We discuss the procedure of adding flag gadgets in \cref{sec:state prep appendix}, and remark that there may be ample room for optimizing this procedure to further reduce the logical error and discard rates of an encoding circuit, which we leave for future work.

\subsubsection{Details of magic state proxy circuits}
\label{sec:Step-by-step proxy simulations}

\begin{figure}
  \centering
  \includegraphics[width=\linewidth]{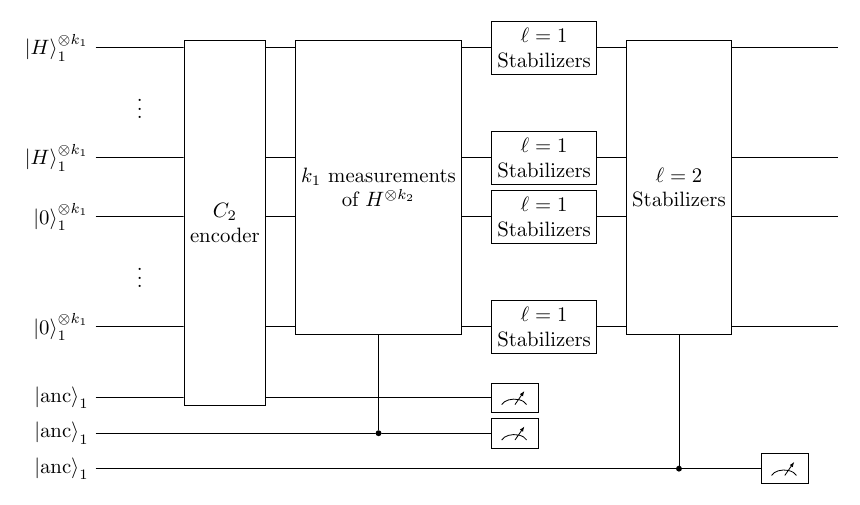}
  \caption{Structure of the concatenated cultivation circuit for $\ket{H}^{\otimes k_1k_2}_{2}$. The input to the circuit is $k_2$ $\ket{H}^{\otimes k_1}_{1}$ states, $n_2-k_2$ $\ket{0}^{\otimes k_1}_{1}$ states, and level-1 ancillas $\ket{\text{anc}}_1$ (as necessary).}
  \label{fig:high level circuit}
\end{figure}

The concatenated cultivation circuits for preparing $\ket{H}_2^{\otimes K}$ with the example codes in \cref{eq:code_A,eq:code_B,eq:code_C} are non-Clifford and too large for direct benchmarking (though recent developments in classically simulating Clifford+$T$ circuits bring the prospect of direct benchmarking closer~\cite{takada2026exact}).
We therefore benchmark concatenated cultivation by simulating the cultivation of a Clifford proxy state, $\ket{+}_2^{\otimes K}$.
These proxy simulations allows us to test the scaling predictions in \cref{sec: noise}, and estimate logical error and discard rates.
Error and post-selection rates are reported subject to the noise model and decoder described in \cref{sec: noise}.
The structure of the encoding circuit is shown \cref{fig:high level circuit}.

The first step is to run the fault-tolerant encoding circuit on all $C_1$ code blocks.
In the examples we analyze, this corresponds to the encoding circuit for the $\params{6,2,2}$ H code shown in \cref{fig:C6 encoder}, but with the $CH$ gates replaced by $CX$ gates.
We then implement the fault-tolerant encoding circuit of $C_2$ using the logical gates of $C_1$.

We then perform a logical operator measurement using a Clifford proxy for the circuit in \cref{eq:meas_h_end}, replacing $CH$ gates by $CX$ gates to measure a logical $X^{\otimes k}$ operator.
Recall from \cref{eq:MH} that performing $CH$ requires two injections of $T_Y$ gates.
Our proxy for these $T_Y$ gates is to sandwich the $CX$ gate in a standard measurement gadget by two trivial (identity-gate) injections on either side, which are stand-ins for the two injections necessary for a $T$ gate (the ``bare'' $T$-state injection and a feed-forward $S$ gate).
In total, our Clifford proxy for each $CH$ gate in \cref{eq:meas_h_end} is the gadget
\begin{align}
\vcenter{\hbox{
\begin{tikzpicture}[scale=0.8, transform shape]
\begin{yquant}
  qubit {$\ket{0}$} q0;
  qubit {data} q1;
  qubit {ancilla} q2;
  cnot q1 | q0;
  [label = $X$] measure q0;
  discard q0;
  init {$\ket{0}$} q0;
  cnot q1 | q0;
  [label = $X$] measure q0;
  discard q0;
  align q0, q1, q2;
  cnot q1 | q2;
  align q0, q1, q2;
  init {$\ket{0}$} q0;
  cnot q1 | q0;
  [label = $X$] measure q0;
  discard q0;
  init {$\ket{0}$} q0;
  cnot q1 | q0;
  [label = $X$] measure q0;
  discard q0;
\end{yquant}
\end{tikzpicture}
}},
\end{align}
A quick comment on the overheads of this gadget.
The example codes $\params{36,4,4},\params{48,8,4}$, and $\params{96,14,6}$ all use the $\params{6,2,2}$ H code for their level-$1$ code blocks, which only requires an $S$ injection half of the time because the feed-forward gates for $T^{\otimes 2}$ are uniformly random on $\set{I,S}^{\otimes 2}$ (and $S^{\otimes 2}$ is transversal).
On average then, executing the level-$2$ Hadamard measurement will require $3n_2$ injections of level-$1$ states: $2n_2$ for the $T$-state injections and $n_2$ (on average) for the feed-forward $S$ gates.
Our proxy simulations restrict to the worst case; every level-$2$ $X^{\otimes k}$ measurement includes $4n_2$ injections of level-$1$ states.
In the results reported below, we do not post-select on the syndrome outcomes of the level-$1$ ancilla code blocks used for the injections, though this is an extra filter that can be incorporated.

The final step of concatenated cultivation is to measure stabilizers of the concatenated code.
We perform one round of level-$1$ stabilizer measurements on each level-$1$ code using level-$0$ (physical) Bell-state ancilla qubits (see \cref{fig: C6 stab} in \cref{sec:additional_circuits}).
For code $\params{96,14,6}$ with $d_2 > 2$, we then measure level-$2$ stabilizers using a level-$1$ Bell-state ancilla code blocks.
In the case of $d_2=2$ (codes $\params{36,4,4}$ and $\params{48,8,4}$) we do not measure level-$2$ stabilizers because doing so does not (asymptotically) reduce the $O(p^2)$ probability of an error that triggers erasure when consuming a cultivated state in a quantum algorithm.

\subsubsection{Results of magic state proxy simulations}

\begin{figure}[t]
  \centering
  \includegraphics{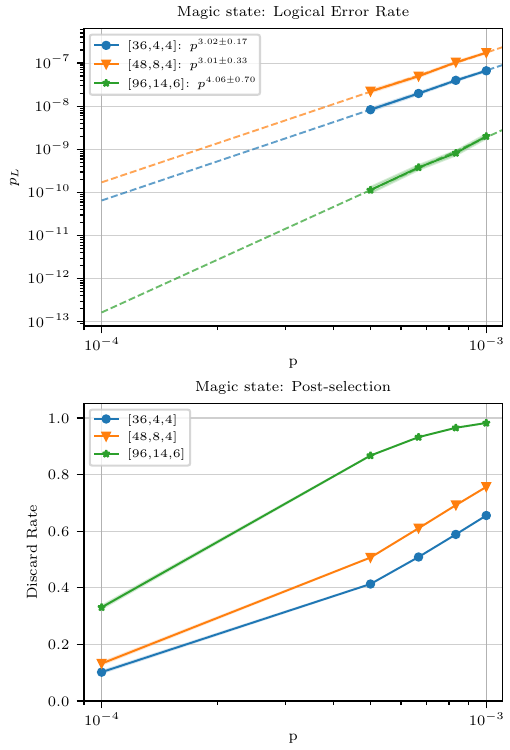}
  \caption{
  Logical error rate $p_L$ and discard rate for the cultivated $\ket{+}^{\otimes K}_2$ state in codes $\params{36,4,4}, \params{48,8,4}$ and $\params{96,14,6}$.
  Shaded regions reflect 95\% confidence intervals.
  }
  \label{fig:magic state plot}
\end{figure}

Our discussions in \cref{sec: decoding comments} and \cref{sec: noise} predict that the logical error rates of our example concatenated cultivation circuits should scale as $O(p^{\min\{4,t+2\}})$.
We therefore expect our distance-4 examples ($\params{36,4,4}$ and $\params{48,8,4}$) to scale as $O(p^3)$, and the distance-6 example ($\params{96,14,6}$) to achieve the full $O(p^4)$ suppression provided by the cultivation protocol.
The results of our proxy simulations, shown in \cref{fig:magic state plot}, support this prediction.

In the first eight columns of \cref{tab:magic_state_resources} we report metrics for the simulated circuits along with the logical error and discard rates at a physical error rate of $p = 5 \times 10^{-4}$.
We emphasize that there are clear avenues to improve the spacetime overheads of these circuits.
Firstly, the total qubit count is an overly conservative estimate here; from  \cref{fig:C6 encoder} we see that $25 \% $ of the physical qubits in the level-1 $\params{6,2,2}$ code blocks are ancillas.
These ancillas can be reused later on in the encoding circuit but this is not done currently.
As another example, the $\ell=1$ and $\ell = 2$ stabilizer measurements are performed serially as shown in \cref{fig: C6 stab}.
More careful optimization of the sub components of \cref{fig:high level circuit} would no doubt increase performance.

\begin{table*}[t]
  \centering
  \caption{Circuit metrics and performance at $p = 5\times 10^{-4}$. $p_L$ and $C = \text{qubits} \circ \text{depth} / (1 - \text{discard})$ are the logical error rate and spacetime cost of concatenated cultivation. The final two columns provide comparison with cultivation on the surface code.
  $p_L^{\text{SC}}=1-(1-p_L)^{1/K}$ is the error rate required of each cultivated surface code patch in order match the error rate of concatenated cultivation.
  $C_\text{SC}$ is the spacetime cost for cultivating $K$ surface code magic states at $p_L^{\text{SC}}$.
   The value of $p_L^{\text{SC}}$ obtained by code $\params{96,14,6}$ is unachievable with the surface-code cultivation methods of Ref.~\cite{gidney2024magicstatecultivationgrowing}, so the corresponding cost $C_\text{SC}$ is bounded by that of cultivating at $p_L^{\text{SC}}=3.6\times10^{-11}$.}
  \begin{tabular}{|c||c|c|c|c||c|c|c||c|c|}
    \hline
     Code &
     Qubits &
     CNOTs &
     Depth &
     Meas. &
     \makecell{Discard \\ rate} &
    \makecell{Error rate $p_L$ \\ (95\% CI)} &
    \makecell{spacetime \\ cost $C$} &
    $p_L^{\text{SC}}$ &
    \makecell{spacetime \\ cost $C_\text{SC}$}\\
    
    \hline
    $\params{36,4,4}$ &  112 & 880 & 73 & 272 & 41.3\% & $\displaystyle 8.29^{+0.64}_{-0.60}\times10^{-9}$  & $1.39 \times 10^{4}$ & $2.07^{+0.16}_{-0.15} \times 10^{-9}$ & $1.61 \times 10^{5}$ \\
    
    $\params{48,8,4}$ &  144 & 1172 & 74 & 356 & 50.7\% & $\displaystyle 2.21^{+0.15}_{-0.14}\times10^{-8}$  & $2.16 \times 10^{4}$ & $2.76^{+0.19}_{-0.18} \times 10^{-9}$ & $3.20 \times 10^{5}$ \\
    
    $\params{96,14,6}$ & 256 & 3142 & 164 & 854 & 86.7\% &  $\displaystyle 1.14^{+0.20}_{-0.17}\times10^{-10}$ & $3.16 \times 10^{5}$ & \textcolor{red}{$8.14^{+1.45}_{-1.23} \times 10^{-12}$} & \textcolor{red}{$\mathbf{>5.98 \times 10^{5}}$} \\
    
    \hline
  \end{tabular}
  \label{tab:magic_state_resources}
\end{table*}

One way to contextualize the performance of concatenated cultivation (\cref{tab:magic_state_resources} and \cref{fig:magic state plot}) is to compare with magic state cultivation in the surface code \cite{gidney2024magicstatecultivationgrowing}.
Specifically, we focus on the spacetime cost for producing a magic state at given logical error rate.
There is a subtlety in the comparison as our method is batched, while cultivation on the surface code allows individual magic states to be produced sequentially.
A given instance of our protocol produces $\ket{H}^{\otimes K}$ with logical error rate $p_L$ and naive spacetime cost $C = \frac{\text{qubits } \circ \text{ depth}}{\text{1 - discard rate}}$.
We compare $C$ with the spacetime cost $C_\text{SC}$ of producing $\ket{H}^{\otimes 
K}$ with logical error rate $p_L$ via surface code cultivation.
To estimate $C_\text{SC}$, we find the spacetime cost $C^{(0)}_\text{SC}$ of cultivating a state with logical error rate $p_L^{\text{SC}} = 1-(1-p_L)^{1/K}$ and set $C_\text{SC} = K C^{(0)}_\text{SC}$.

Using the noise model described in \cref{sec: noise} we simulate surface code cultivation at $p=5 \times 10^{-4}$ and toggle the complimentary gap to achieve the target single qubit error rate $p_L^{\text{SC}}$, which gives a value for $C^{(0)}_\text{SC}$.
Our simulations consist of performing two rounds of cultivation at $d=3$ and $d=5$ and then embedding the $d=5$ color code code into $d=11$ surface code.
Ref.~\cite{gidney2024magicstatecultivationgrowing} reports spacetime cost in $\text{qubit} \circ \text{rounds}$ where each round roughly corresponds to a depth-10 circuit.
We note that the estimate of this cost incorporates that not all qubits in the final $d=11$ patch are used throughout the circuit, which is not done for the concatenated cultivation circuits.
Thus the comparison of the spacetime cost should be understood to be generous to the surface code cultivation circuit.

In the final three columns of \cref{tab:magic_state_resources} we report the spacetime cost $C$, the corresponding surface-code-cultivated logical error rate $p_L^{\text{SC}}$ and the spacetime cost $C_\text{SC}$.
In the case of code $\params{96,14,6}$, a logical error rate of $p_L^{\text{SC}} \sim 10^{-12}$ is not achievable and likely requires an additional level of cultivation
at $d=7$.
In place, of this we used the minimal value of $p_L^{\text{SC}} = 3.6\times 10^{-11}$ obtained by cultivation at $d=5$ in \cref{tab:magic_state_resources}, and simply note that the reported value for $C_\text{SC}$ is a significant underestimate. Nevertheless, in all cases the associated spacetime cost for cultivation $C_\text{SC}$ is more than one order of magnitude larger than the cost $C$ of the corresponding concatenated protocol.

\subsection{CSS resource states for Clifford gates}

\begin{figure}
    \centering
    \includegraphics[width=\linewidth]{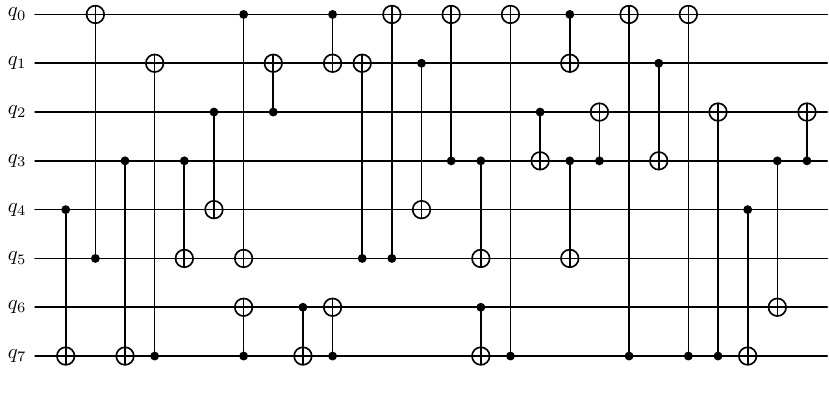}
    \caption{Logical circuit implemented by the $CX$-type resource state benchmarked for code $\params{48,8,4}$.}
    \label{fig:C8 CNOT main}
\end{figure}

In this section, we report benchmarks for fault-tolerantly preparing the CSS states need to perform logical Clifford gates via injection.
Recall from \cref{thm:css_clifford} that the resource states for $S/H$-type gates are tensor products of logical $\ket{0}$ and $\ket{+}$ states on a single $\params{N,K,D}$ block.
The representative $S/H$-type resource states that we compile for our benchmarks correspond to the logical gates 
\begin{align*}
    \params{36,4,4}: &~S_1S_2 \\
    \params{48,8,4}: &~S_1S_2S_5S_7\\
    \params{96,14,6}: &~S_1S_2S_5S_7S_{10}S_{11}S_{12}S_{13}.
\end{align*}
$CX$-type resource states are injected via Knill-QEC teleportation and require two $\params{N,K,D}$ blocks, as indicated in \cref{fig:main}(e).
For the $CX$ gate injection, we benchmark $CX$-type resource states for circuits with 9, 30, and 74 logical $CX$ gates, respectively, for codes $\params{36,4,4}$, $\params{48,8,4}$ and $\params{96,14,6}$.
The logical $CX$ circuit for code $\params{48,8,4}$ is shown in \cref{fig:C8 CNOT main}, and the remaining circuits are provided in \cref{fig:CNOTs}.

All Clifford resource-state circuits are obtained using the flag-at-origin method in Ref.~\cite{forlivesi2025flag}, which we also briefly describe in \cref{sec: flag at origin}.
We additionally perform a search over 1000 randomly sampled bipartite graph states that are equivalent to the CSS state that we want to prepare.
Flag-at-origin adds the flag gadgets to each of of these circuits and selects the one that presents the smallest total probability of a physical fault across the fault locations in the resulting circuit.
For each resource state, we prepare two encoding circuits that respectively minimize either qubit count (space) and circuit depth (time).
Below, we omit the data for the space-optimized circuit for code $\params{96,14,6}$ because the discard rate was nearly 100\% for values of $p \geq 5\times 10^{-4}$.

\begin{figure*}
    \centering
    \subfigure[$S/H$ state.\label{fig:decoded_results_SH}]{%
        \includegraphics[width=0.48\textwidth]{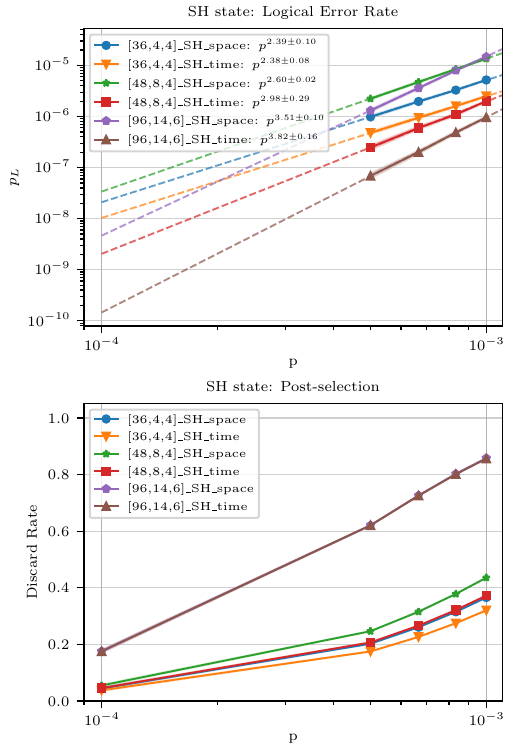}%
    }\hfill
    \subfigure[CNOT state.\label{fig:decoded_results_CNOT}]{%
        \includegraphics[width=0.48\textwidth]{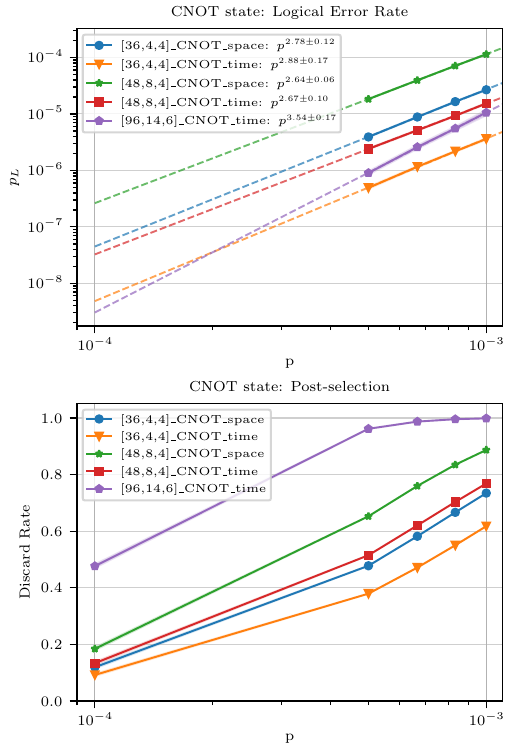}%
    }
    \caption{
    Logical error rates $p_L$ (top) and post-selection discard rates (bottom) for the flag-at-origin resource states.
    Dashed lines in logical error rate plots show power-law fits, the fitted exponent is included in the legend.
    Shaded regions reflect 95\% confidence intervals.
    }
    \label{fig:Clifford decoded results}
\end{figure*}

\cref{fig:Clifford decoded results} shows that the circuits minimizing depth outperform those minimizing space both in logical error rate and discard rate.
This is expected due to inclusion of idling error in our noise model.
We find that the tradeoffs between space and time contribute unequally to the overall spacetime volume, with the depth-optimized circuits having much smaller spacetime volume than the space-optimized circuits.

\begin{table*}
  \centering
  \caption{Metrics for our CSS resource states found via flag-at-origin. The first column reports the minimum factory size for the distillation approach \cite{zheng2020constant} which can be compared to the total number of qubits used in the flag-at-origin circuits in column two. The final two columns show discard and logical error rates at $p=5\times10^{-4}$. Consuming the CX states implements the logical circuits shown in \cref{fig:CNOTs}.}
  \label{tab:resource_table}
  \renewcommand{\arraystretch}{1.3} %
  \begin{tabular}{|c||c||c|c|c|c||c|c|c|}
    \hline
    \makecell{ Circuit } &
    \makecell{Distillation \\ factory size \\ (qubits) \cite{zheng2020constant}} &
    Qubits &
    Depth &
    \makecell{CNOTs} &
    Meas &
    \makecell{Discard \\ Rate} &
    \makecell{Error Rate $p_L$ \\ (95\% CI) } \\
    \hline
    \hline 
$\params{36,4,4}$\_CNOT\_time & 3600 & 264 & $\bm{48}$ & 700 & 192 & $\bm{37.93\%}$ & $\bm{4.92^{+0.18}_{-0.17}\times10^{-7}}$ \\
$\params{36,4,4}$\_CNOT\_space & 3600 & $\bm{130}$ & 569 & 700 & 192 & 47.78\% & $3.92^{+0.05}_{-0.05}\times10^{-6}$ \\
\hline
$\params{48,8,4}$\_CNOT\_time & 4800 & 378 & $\bm{64}$ & 1046 & 282 & $\bm{51.50\%}$ & $\bm{2.41^{+0.04}_{-0.04}\times10^{-6}}$ \\
$\params{48,8,4}$\_CNOT\_space & 4800 & $\bm{180}$ & 811 & $\bm{1022}$ & $\bm{274}$ & 65.27\% & $1.83^{+0.01}_{-0.01}\times10^{-5}$ \\
\hline
$\params{96,14,6}$\_CNOT\_time & 17640 & 1415 & 186 & 3985 & 1235 & 96.19\% & $9.01^{+0.57}_{-0.53}\times10^{-7}$ \\

\hline

$\params{36,4,4}$\_SH\_time & 900 & 120 & $\bm{26}$ & 282 & 84 & $\bm{17.50\%}$ & $\bm{4.79^{+0.35}_{-0.32}\times10^{-7}}$  \\
$\params{36,4,4}$\_SH\_space & 900 & $\bm{65}$ & 244 & 282 & 84 & 20.23\% & $9.83^{+0.50}_{-0.47}\times10^{-7}$  \\
\hline
$\params{48,8,4}$\_SH\_time & 1200 & 148 & $\bm{28}$ & $\bm{344}$ & $\bm{100}$ & $\bm{20.69\%}$ & $\bm{2.48^{+0.26}_{-0.23}\times10^{-7}}$  \\
$\params{48,8,4}$\_SH\_space & 1200 & $\bm{77}$ & 292 & 348 & 102 & 24.66\% & $2.23^{+0.07}_{-0.07}\times10^{-6}$  \\
\hline
$\params{96,14,6}$\_SH\_time & 4410 & 495 & $\bm{91}$ & 1266 & 405 & 62.05\% & $\bm{6.90^{+0.81}_{-0.72}\times10^{-8}}$  \\
$\params{96,14,6}$\_SH\_space & 4410 & $\bm{169}$ & 802 & $\bm{994}$ & $\bm{264}$ & 62.07\% & $1.31^{+0.08}_{-0.08}\times10^{-6}$  \\

\hline
  \end{tabular}
  \label{tab: clifford metrics}
\end{table*}

As discussed in \cref{sec:Clifford}, the alternative method for producing resource states for Clifford gates, distillation, has the drawback that the minimum factory size for producing a single copy of a resource state is large: $n(d+1)^2$.
This was one of our primary motivations for employing the flag-at-origin method, which achieves the same fault distance but has a significantly smaller minimum factory size.
The qubit requirements for both methods are reported in \cref{tab:resource_table}.

We leave a more granular analysis of the overheads for specific codes and error models---factoring in spacetime cost and acceptance rates---for future work.
While flag-at-origin circuits achieve the expected scaling of logical error rate, we note that the values of $p_L$ are at least an order of magnitude worse than the their counterparts for the proxy magic state distillation in \cref{sec:Step-by-step proxy simulations}.
This discrepancy suggests that there may be better methods for fault tolerant preparation of CSS states in these codes that leverage the iterative structure of concatenation.
Indeed, flag-at-origin is an algorithmic method for producing fault tolerant circuits that prepare arbitrary CSS states but it makes no claims about optimality. 

\section{Discussion}
\label{sec: discussion}

We have presented a scheme for universal fault-tolerant quantum computation in which almost every logical gate is injected via teleportation, consuming fault-tolerantly prepared resource states.
Arbitrary multi-qubit Clifford circuits require at most 9 CSS resource states, and a universal gate set is completed by consuming logical magic states.
As a by-product, injecting these resource states yields full (Knill) or partial (Steane) syndrome measurements, reducing the need for additional gadgets used purely for syndrome extraction.
Our scheme is restricted to the family of concatenated codes built from SWEL seed codes because it is precisely for this family that both classes of resource state can be prepared with high fidelity and low overhead.
The transversal global $H^{\otimes k}$ and $S^{\otimes k}$ gates allow us to reduce every Clifford to CSS resource states preparable by flag-at-origin, and enables the concatenated cultivation protocol.

The numerical results in this paper provide an initial benchmarking of the quality of the CSS and magic state resource states.
This is a necessary first step, but it is deliberately local in scope: we have characterized the cost per resource state rather than the cost of a computation.
Our results suggest that logical circuits in the megaquop and gigaquop regimes are executable on codes and at physical error rates that near term hardware can accommodate, but a study with a more global view is a natural next step.
Concretely, one would like end-to-end estimates of the cost of executing a Bruhat-decomposed Clifford circuit---accounting for the $O(1)$ injections per stage and the online rejection incurred at each injection---as well as the cost of synthesizing single-qubit rotations to a target precision $\epsilon$, where the $T$-count and hence the number of $\ket{H}^{\otimes k}$ states scales as $O(\log(1/\epsilon))$.
Combining our per-state benchmarks with a suitable compiler would yield spacetime-cost estimates for complete algorithms and allow a fair comparison against other high-rate computing architectures.
We provide further discussion about compilation, including a scheme to minimize latency in $T$-gate layers and a scheduler to minimize the $T$-depth of a circuit, in \cref{sec: Loose ends}.

As mentioned in \cref{sec:examples}, there are clear pathways to improving the resource state preparation circuits.
When scaling up these codes, both by increasing the size of the seed codes and the level $L$ of concatenation, the lookup tables used for decoding in this work will not be viable.
In this case algorithmic decoders that incorporate concatenation structure can be used to achieve good performance~\cite{goto2024highperformance}.
On another note, in this work we have imposed homogeneity between the code blocks containing data qubits and code blocks storing resource states, but relaxing this distinction should decrease overheads and flexibility of compilation.
For this to be possible, there needs to be compatible structure existing between the data block and the ancilla block, such as a homomorphic logical CNOT~\cite{benhemou2026automatedlogicalcliffordgadgets}.
In this case, gate injections for certain resource states can in principle be performed.
For concatenated codes, computationally tractable searches for these homomorphic logical CNOT can be carried out on seed codes and then chained together to yield fault tolerant inter-block logical CNOT gates for large distance codes.

\section{Acknowledgments}

We acknowledge helpful correspondence with Shival Dasu and Hayato Goto.
We thank Ruslan Shaydulin and Rob Otter for executive support of this work.

\bibliography{main.bib}

\appendix
\onecolumngrid

\section{Seed code definitions}
\label{sec: Code Definitions}

Here we define the seed codes for the level-$2$ concatenated codes in the main text.
All of the seed codes are SWEL, with identical support for their $X$ and $Z$-type stabilizes and logical Pauli operators.
The matrices whose rows define the support of $X$-type stabilizers ($S_X$), $Z$-type stabilizers ($S_Z$), $X$-type logical single-qubit Pauli operators  ($L_X$), and $Z$-type logical single-qubit Pauli operators ($L_Z$) are as follows.
\begin{itemize}
\item[]
$\params{6,2,2}$:
\begin{align}
  S_X = S_Z =
  \begin{pmatrix}
    1 & 1 & 1 & 1 & 0 & 0 \\
    0 & 0 & 1 & 1 & 1 & 1
  \end{pmatrix},
  &&
  L_X = L_Z =
  \begin{pmatrix}
    1 & 0 & 1 & 0 & 1 & 0 \\
    0 & 1 & 0 & 1 & 0 & 1
  \end{pmatrix}.
\end{align}
\item[]
$\params{8,4,2}:$
\begin{align}
  S_X = S_Z =
  \begin{pmatrix}
    1 & 1 & 1 & 1 & 0 & 0 & 0 & 0 \\
    1 & 1 & 0 & 0 & 1 & 1 & 1 & 1
  \end{pmatrix},
  &&
  L_X = L_Z =
  \begin{pmatrix}
    0 & 1 & 0 & 1 & 1 & 0 & 0 & 0 \\
    0 & 1 & 0 & 1 & 0 & 1 & 0 & 0 \\
    0 & 1 & 0 & 1 & 0 & 0 & 1 & 0 \\
    0 & 1 & 0 & 1 & 0 & 0 & 0 & 1
  \end{pmatrix}.
\end{align}
\item[]
$\params{15,7,3}:$
\begin{align}
  S_X = S_Z =
  \begin{pmatrix}
    0 & 0 & 0 & 0 & 0 & 0 & 0 & 1 & 1 & 1 & 1 & 1 & 1 & 1 & 1 \\
    0 & 0 & 0 & 1 & 1 & 1 & 1 & 0 & 0 & 0 & 0 & 1 & 1 & 1 & 1 \\
    0 & 1 & 1 & 0 & 0 & 1 & 1 & 0 & 0 & 1 & 1 & 0 & 0 & 1 & 1 \\
    1 & 0 & 1 & 0 & 1 & 0 & 1 & 0 & 1 & 0 & 1 & 0 & 1 & 0 & 1
  \end{pmatrix},
  &&
  L_X = L_Z =
  \begin{pmatrix}
    1 & 1 & 0 & 1 & 0 & 0 & 0 & 1 & 0 & 0 & 0 & 0 & 0 & 0 & 1 \\
    1 & 1 & 0 & 0 & 1 & 0 & 0 & 0 & 0 & 1 & 0 & 1 & 0 & 0 & 0 \\
    1 & 1 & 0 & 0 & 0 & 1 & 0 & 0 & 0 & 0 & 1 & 0 & 0 & 1 & 0 \\
    1 & 1 & 0 & 0 & 0 & 0 & 1 & 0 & 1 & 0 & 0 & 0 & 1 & 0 & 0 \\
    1 & 0 & 0 & 1 & 0 & 1 & 0 & 0 & 1 & 1 & 0 & 0 & 0 & 0 & 0 \\
    1 & 0 & 0 & 1 & 0 & 0 & 1 & 0 & 0 & 0 & 0 & 1 & 0 & 1 & 0 \\
    1 & 0 & 0 & 0 & 0 & 0 & 0 & 1 & 0 & 1 & 0 & 0 & 1 & 1 & 0
  \end{pmatrix}.
\end{align}
\end{itemize}

\section{Concatenated H codes}
\label{sec: H codes}

\begin{figure}
    \centering
    \includegraphics[width=0.8\linewidth]{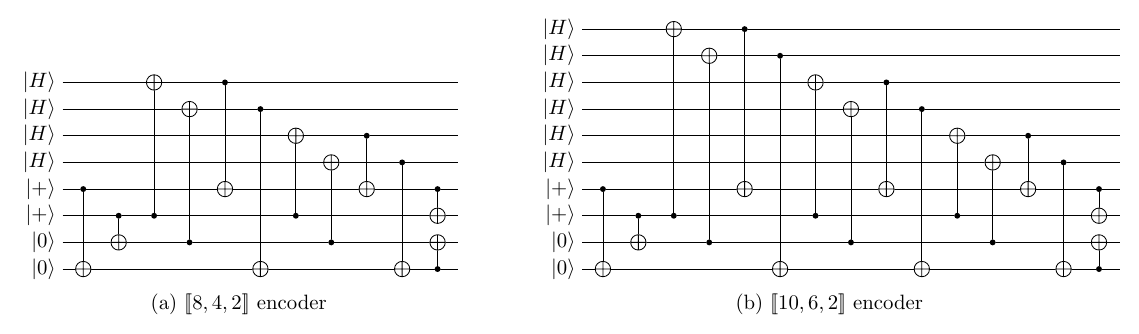}
    \caption{Examples of H code encoders. The encoders are general. A Hadamard measurement gadget is required for FT preparation of $\ket{H}^{\otimes k}$.}
    \label{fig:C8 C10 encoder}
\end{figure}

Here we present some further exploration of concatenated cultivation with $\params{k+4,k,2}$ H codes~\cite{eczoo_quantum_h, Jones_2013}.
As discussed in the main text, the encoding circuit of the full concatenated code $\params{N,K,D}$ relies on recursive use of FT encoders of the seed codes.
Fortunately, encoding circuits for the H codes have a simple structure.
At the input to the H code encoding circuits, we choose the physical qubits in $[0,k-1]$ to contain the logical information, and the pairs $[k,k+1]$ (initialized in $\ket{++}$) \& $[k-2,k-3]$ (initialized in $\ket{00}$) to evolve to $X$ \& $Z$ stabilizers.
Using the format $CX ~ [c_0, t_0] ~ [c_1, t_1] \hdots $ to represent a sequence of $CX$ gates between control ($c_j$) and target ($t_j$) qubits, the encoding circuit is
\begin{align*}
  \begin{tabular}{lll}
    $CX$ & $[k, k+3]$ $[k+1, k+2]$ & \\
    $CX$  & $[k+1,2q]$ $[k+2,2q+1]$ $[2q,k]$ $[2q+1,k+3]$ &(repeat over pairs $[2q,2q+1]$ with $q<k/2$) \\
    $CX$ & $[k,k+1]$ $[k+3,k+2]$ &
  \end{tabular}.
\end{align*}
We provide example circuits in \cref{fig:C6 encoder,fig:C8 C10 encoder}.

\begin{figure}
    \centering
    \includegraphics[width=\linewidth]{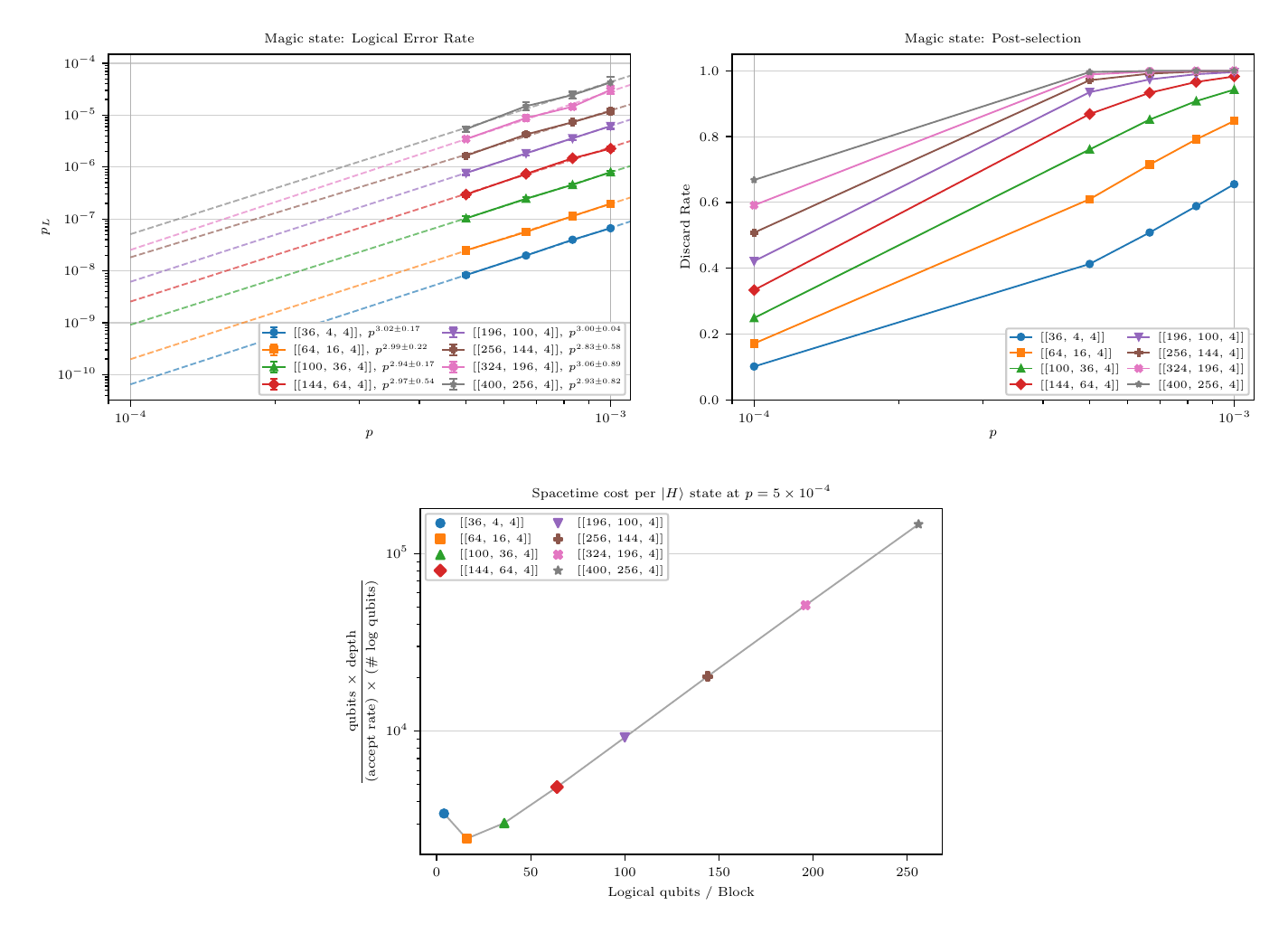}
    \caption{Logical error rate, Discard Rate and Spacetime cost per H state of the self concatenated H codes $\params{(k+4)^2,k^2,4}$.}
    \label{fig:H code benchmarks}
\end{figure}

A nice property of this encoding circuit is that it satisfies \cref{cond:detection} without needing additional flag gadgets (unlike the $\params{15,1,3}$ Hamming example in \cref{fig:hamming encoder}).
We prepare the concatenated cultivation circuits for self-concatenated H codes at level-$2$, $\params{(k+4)^2,k^2,4}$ for $k \in [2,4,6,\hdots , 16]$, by porting these encoding circuits into \cref{fig:high level circuit}.
The benchmarking results are shown in \cref{fig:H code benchmarks}, where in addition to logical error rates and discard rates we plot the spacetime cost per $\ket{H}$ state at $p=5\times 10^{-4}$.
This quantity is given by (qubit number of encoder)$\times$ (encoder depth)/(acceptance rate $\times$ $k$) and is a rough proxy for the quantum resources involved to produce one fault tolerant magic state.
Note this quantity is normalized by the number of logical qubits per block, so \cref{fig:H code benchmarks} points to the $\params{64,16,4}$ code being most efficient for creating magic states at $p=5\times 10^{-4}$.

\section{Fault tolerant circuits for magic states}

\label{sec:state prep appendix}

Here we provide more details on some of the subcircuits and flag gadgets used the concatenated cultivation circuits.
The protocol is shown schematically in \cref{fig:high level circuit}.

\subsection{Flag gadgets for individual undetectable faults}

Concatenated cultivation relies on encoding circuits that posses no individual Pauli faults violating \cref{cond:detection}.
Here we describe two types of gadgets that we use to modify generic encoding circuits to catch the problematic errors that violate \cref{cond:detection}.

Our strategy for producing suitable encoding circuits is as follows.
We first use the Munich Quantum Toolkit (MQT)~\cite{mqt} to produce a general encoding circuit.
We then enumerate all individual faults that are equivalent to pure-logical errors and violate \cref{cond:detection}, identify those faults that are caught by the ``simple'' $X/Z$ flag gadgets described in \cref{eq:simple_flags}, and add the corresponding flag gadgets to the circuit accordingly.
Finally, we identify a minimal set of logical Pauli operators that anti-commute with the remaining problematic errors, and sandwich the encoding circuit with corresponding spacetime flags described \cref{eq:spacetime_flags}.

\subsubsection{Simple $X/Z$ flag gadgets}
\label{eq:simple_flags}

The basic idea for the simple $X/Z$ flag gadgets is to first identify ``star''-type subcircuits of $CX$ gates, which have a shared ``central'' qubit as either the control or the target of all $CX$ gates.
These subcircuits are compatible with gadgets to catch any multi-qubit Pauli error that, when propagated to the end of the subcircuit, addresses the central qubit by an $X$ or $Z$ error (for the $X$ and $Z$ gadgets, respectively).
Specifically, the $X$ and $Z$ flag gadgets are, respectively,
\begin{align}
    \includegraphics[valign=c]{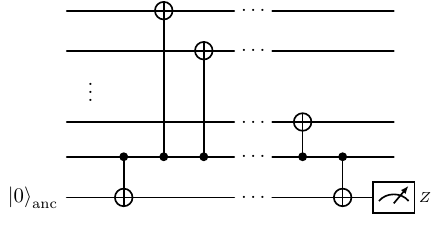}
    &&
    \includegraphics[valign=c]{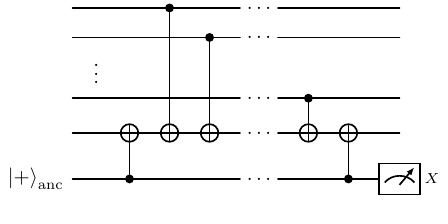}.
\end{align}

\subsubsection{Spacetime gadgets}
\label{eq:spacetime_flags}

Suppose a single fault in an encoding circuit generates a pure logical error that anti-commutes with a logical Pauli $P$ on logical qubit $q$, denoted $\bar P_q$.
We can catch this fault by entangling physical qubit $q$ with an ancilla before running the encoding circuit, and then disentangling logical qubit $q$ the ancilla after the running the decoding circuit~\cite{Chao_2018}.
In the case of $q=0$, this gadget looks like
\begin{align}
  \includegraphics[valign=c]{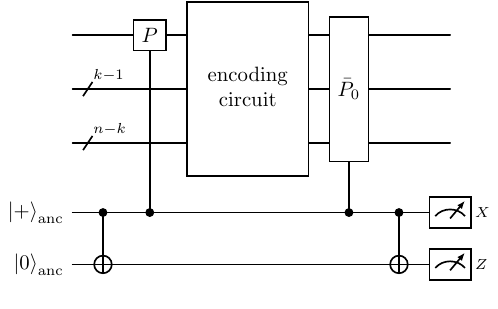}.
\end{align}

\subsection{Additional circuits}
\label{sec:additional_circuits}

\begin{figure}[H]
  \centering
  \includegraphics[width=.9\linewidth]{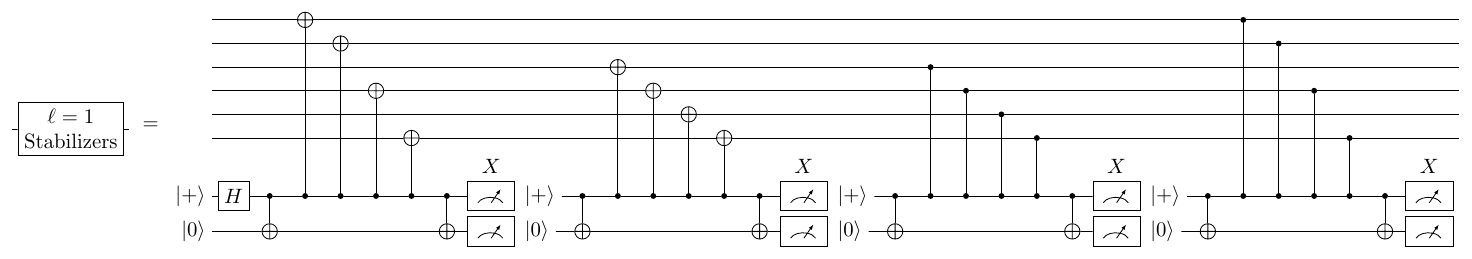}
  \caption{Stabilizer measurements used in \cref{fig:high level circuit}.}
  \label{fig: C6 stab}
\end{figure}

\begin{figure}[H]
  \centering
  \includegraphics[width=0.5\linewidth]{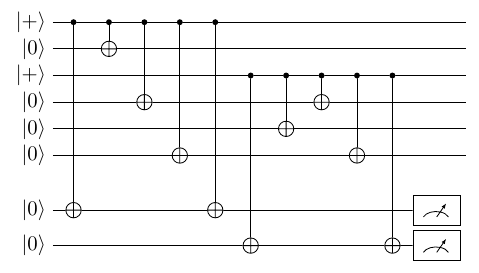}
  \caption{Encoding circuit for the $\ket{00}$ state of the $\params{6,2,2}$ H code.}
  \label{fig:zero state prep}
\end{figure}

Here we include some additional circuits used in our protocol. 
We use the measurement circuit in \cref{fig: C6 stab} for the $\ell = 1$ stabilizers.
The $\ell = 2$ stabilizer measurements are also measured in this serial fashion.
As mentioned in \cref{sec:non-clifford}, we make no effort to optimize these circuits or to identify smaller subsets of the stabilizers sufficient to catch logical errors. 
When encoding $C_1$ code blocks into the $C_2$ code, ``ancilla'' $C_1$ code blocks whose logical Pauli operators get mapped to stabilizers of $C_2$ are encoding using the circuit in \cref{fig:zero state prep}, rather than the general encoding circuit in \cref{fig:C6 encoder}.


\section{CSS states}

This appendix is a supplement to \cref{sec:Clifford}.

\subsection{Flag at origin}
\label{sec: flag at origin}

Fault-tolerant preparation of logical states is a prerequisite for fault-tolerant quantum computation. In order to leverage the distance $d$ of a code, the preparation circuits of logical states of that code must have circuit fault distance of $d$. This means ensuring that any combination of $f \leq t = \lfloor(D-1)/2\rfloor$ faults that propagates to an error of minimum weight $w \leq f$ up to stabilizers is detectable. Prior approaches are limited by poor scalability (flag-based methods) or high overhead from repeated verification rounds. In the case of CSS states, which is all that is required to perform Clifford gates in our protocol, the flag-at-origin protocol~\cite{forlivesi2025flag} provides a modular, single-shot construction.

The protocol exploits a structural property of CSS state-preparation circuits: any CSS state can be prepared using a bipartite $CX$ circuit in which qubits are partitioned into sets $A$ and $B$, with all $CX$ directed from $A$ (controls) to $B$ (targets). For circuits with this bipartite structure $X$ and $Z$ errors can be handled separately:
\begin{itemize}
  \item $X$-type errors propagate \emph{forward}: an $X$ fault on a control qubit spreads $X$ errors onto target qubits in subsequent layers.
  \item $Z$-type errors propagate \emph{backward}: a $Z$ fault on a target qubit effectively infects earlier control qubits.
\end{itemize}
The key idea is to place flag gadgets at the \emph{origin} of each propagation direction---catching errors before they spread out into high-weight, uncorrectable configurations.

\subsection{$CX$ gates}
\label{sec:$CX$ gates}

\begin{figure}[t]
  \centering
  ~\hfill
  \subfigure[$CX$-type gate for $\params{36,4,4}$\label{fig:C6 CNOT}]{%
    \includegraphics[width=0.25\textwidth]{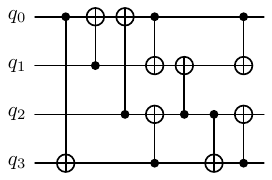}%
  }
  \hfill
  \subfigure[$CX$-type gate for $\params{48,8,4}$\label{fig:C8 CNOT}]{%
    \includegraphics[width=0.4\textwidth]{cnot_circuit8.pdf}%
  }
  \hfill~
  \\
  \subfigure[$CX$-type gate for $\params{96,14,6}$\label{fig:Hamming CNOT}]{%
    \includegraphics[width=0.8\textwidth]{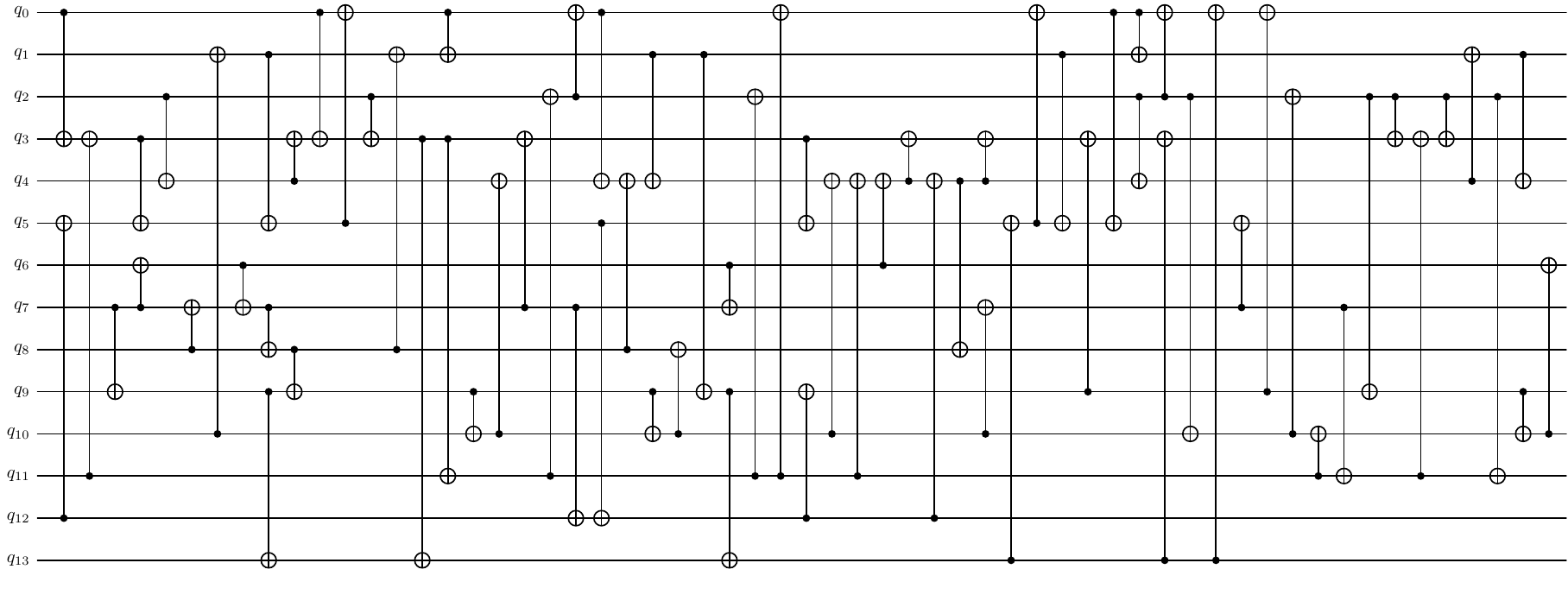}%
  }
  \caption{$CX$-type gate injected by the resource states benchmarked for codes $\params{36,4,4}$ (a), $\params{48,8,4}$ (b) and $\params{96,14,6}$ (c).}
  \label{fig:CNOTs}
\end{figure}

Here we show the exact logical circuits implemented by consuming the $CX$ states prepared in codes $\params{36,4,4}, \params{48,8,4}$ and $\params{96,14,6}$.
For each code we randomly generated a sequence of $\sim K^2/\log_2 K$ $CX$ gates on $K$ qubits.
Specifically, these circuits consist of 9, 30 and 74 logical $CX$ gates for codes $\params{36,4,4}$, $\params{48,8,4}$ and $\params{96,14,6}$; see \cref{fig:CNOTs}.
The stabilizers of the corresponding resource state are obtained by preparing two code blocks into $K$ logical Bell pairs, and applying the $CX$ circuit to one of the code blocks, as in the boxed subcircuit for $\ket{\Phi,U}$ in \cref{eq:gate_telep}.

\section{Loose ends}
\label{sec: Loose ends}
\subsection{Feedforward gates}
\label{sec:feedforward}

The main compilation bottleneck for our protocol is the Clifford correction required after a $T$ injection.
This is common to all magic-state protocols but is exacerbated by the high rate of our code blocks: the measurement outcomes $\bm s \in \mathbb{Z}_2^k$ in \cref{eq:TYI} call for an addressable correction $S^{\otimes\bm s}$, and because $\bm s$ is known only at runtime the corresponding resource state cannot be prepared in advance, while the $2^k$ possible outcomes rule out precompiling them all. One option is simply to absorb this latency, consuming prepared Knill/Steane-QEC resource states---and running a dedicated factory of logical Bell states for Knill-QEC error correction---while the $S$ resource state is made.

To limit this latency one could attempt to precompile as much of the $S$ resource state ahead of time as possible. One way to attack this is to prepare an $\bm s$-independent reference state offline, say $\ket{0}^{\otimes k}$, and shape it at runtime.
First, note that having a free global $S^{\otimes k}$ and free logical Pauli-$Z$ strings means we only need to address at most $\lfloor k/2 \rfloor$ qubits: if $\bm s\in\set{0,1}^k$ is a bitstring and $\bar{\bm s}$ is its bitwise complement, then we can freely convert an $S$ gate on the $\abs{\bm s}$ qubits indicated by $\bm s$ into an $S$ gate on the $\abs{\bar{\bm s}} = k - \abs{\bm s}$ qubits indicated by $\bar{\bm s}$:
\begin{align}
  S^{\otimes\bm s}
  = S^{\otimes k} (S^\dag)^{\otimes \bar{\bm s}}
  = S^{\otimes k} Z^{\otimes\bar{\bm s}} S^{\otimes\bar{\bm s}}.
\end{align}
To produce the required state $H^{\otimes\bm  s}\ket{0}$ we start from the reference state $\ket{0}^{\otimes k}$ and convert $\ket{0} \to \ket{+}$ on the addressed qubits by measuring the corresponding logical $\bar X_i$ operators, applying a $\bar Z_i$ correction on the $-1$ outcome.
A Steane-style measurement would be single-shot but circular---it requires an ancilla in exactly the state $H^{\otimes\bm s}\ket{0}$ we are trying to make---so we can instead perform a Shor-style measurement of each $\bar X_i$ with a cat-state ancilla, verified by post-selected $ZZ$-parity checks.
For each addressed operator the first measurement is the (random) projective state preparation, and the remaining rounds merely confirm its outcome against readout faults; discarding on any disagreement, $\lceil (d+1)/2 \rceil$ rounds suffice for fault distance $d$ (one projection plus $\lfloor (d-1)/2 \rfloor$ confirmations). Each round of logical measurements needs to be followed by a QEC round for the $X$-sector, which can be performed in $O(1)$ depth using an ancillary $\ket{+}^{\otimes k}$ state that can be prepared in advance.

We can tally the space and online time cost of preparing the $S$ resource state via precompilation under the assumption that the $\lfloor k/2 \rfloor \lceil (d+1)/2 \rceil$ verified cat states, $\lceil (d+1)/2 \rceil$ $\ket{+}^{\otimes k}$ states and the $\ket{0}^{\otimes k}$ state are prepared in advanced and available on demand.
The depth of each logical measurement round can be upper bounded by $\lfloor k/2 \rfloor$.
In this case the conversion from $\ket{0}^{\otimes k} \to H^{\otimes\bm s}\ket{0}^{\otimes k}$ is depth $\lfloor k/2 \rfloor\lceil (d+1)/2 \rceil + \lceil (d+1)/2 \rceil$.
To tally the space cost we first note that without additional optimization the cat states use $|X| =\max(|X_i|)_{i \leq k} + 1$ qubits, where $|X_i|$ is the Hamming weight of a representative of the $i$-th $X$ logical operator.
Let $C$ be the qubit cost for making  $\ket{+}^{\otimes k}$ or $\ket{0}^{\otimes k}$.
Then creating the $S$ resource state via conversion costs $\lfloor k/2 \rfloor \lceil (d+1)/2 \rceil|X| +  (\lceil (d+1)/2 \rceil + 1)C$. 

If we use flag-at-origin for the no-precompilation strategy we can compare the $S/H$ depth results in \cref{tab: clifford metrics} to an upper bound for the online depth with the precompilation strategy, $\lfloor k/2 \rfloor\lceil (d+1)/2 \rceil + \lceil (d+1)/2 \rceil$.
Comparisons of qubit cost are murkier because both flag-at-origin and concatenated cultivation can be used to produce the $\ket{+}^{\otimes k}$ and $\ket{0}^{\otimes k}$ states. \cref{tab:magic_state_resources,tab: clifford metrics} provide estimates for the qubit costs $C$ of these approaches, though in cultivation case of \cref{tab:magic_state_resources}, the qubit (along with depth and rejection rate) numbers are large over-estimates because the injection circuits contain a lot of machinery for measuring logical $H$ operators in \cref{fig:high level circuit} (such as logical $T_Y$ gate injections) that is not necessary to prepare $\ket{+}^{\otimes k}$ and $\ket{0}^{\otimes k}$ states.
\cref{tab: precompilation comparison} shows the ratios of online depth and qubit cost for precompilation using concatenated cultivation (with the aforementioned overheads eliminated; replacing \cref{tab:magic_state_resources}) versus time-optimized flag-at-origin circuits (identical to \cref{tab: clifford metrics}).

\begin{table}[b]
  \centering
  \caption{Ratio metrics for precompilation versus no-precompilation. We use concatenated cultivation for precompilation circuits and remove the excess qubits associated with injections that are only needed for $\ket{H}$ states. Logical operators in \cref{sec: Code Definitions} provide conservative estimates of $|X| = 10$, $10$, and $16$ qubits for the $\params{36,4,4}, \params{48,8,4}$, and $\params{96,14,6}$ cat states.}
  \begin{tabular}{|c||c|c|}
    \hline
    Code & \makecell{Online depth \\ ratio} & \makecell{Qubit cost \\ ratio} \\
    \hline
    $\params{36,4,4}$ & 0.346 & 2.767 \\
    $\params{48,8,4}$ & 0.536 & 3.081 \\
    $\params{96,14,6}$ & 0.352 & 2.319 \\
    \hline
  \end{tabular}
  \label{tab: precompilation comparison}
\end{table}

For the precompiled approach the reduction in online depth is roughly commensurate with the increase in qubit cost.
Here qubit cost only accounts for the minimum number of qubits needed to prepare the $S$ resource state and idling of the data qubits during this preparation necessitates a number of QEC cycles -- and therefore Knill/Steane resource states -- extensive in the wait time.
The upshot is that the precompilation strategy likely also has reduced qubit overhead when one looks at the total cost of implementing the $T$ gate fault-tolerantly.
It should be emphasized that this comparison is somewhat simplistic in that it does not incorporate the rejection rates of the ancilla states, which need to be factored in to estimate the expected qubit cost of minimizing data qubit idling.
At the level of the state preparation circuits themselves, methods from Ref.~\cite{goto2024highperformance} point to there being less costly ways to prepare $\ket{0}^{\otimes k}$ and $\ket{+}^{\otimes k}$ states in concatenated codes.
We leave optimizing the compilation of $T$ gate layers in these codes to future work.

\subsection{Compilation and scheduling}
\label{sec:compilation}

Here we briefly discuss the compilation of circuits into the primitives defined in \cref{sec:Clifford} and \cref{sec:non-clifford}.
To this end, we first decompose any quantum circuit into a Clifford+$T$ gate set.
We then repeatedly \emph{sift} the circuit for $T$ gates using the sifting procedure of Ref.~\cite[Algorithm~1]{nottingham2024circuit}.
For a fixed Clifford+$T$ circuit $C$, sifting for $T$ gates partitions $C$ into three sub-circuits, $(C_{\text{before}}, C_T, C_{\text{after}})$ with the following guarantees:
\begin{enumerate}
  \item Applying the sub-circuits in sequence recovers the original circuit, $C = C_{\text{after}} C_T C_{\text{before}}$.
  \item $C_{\text{before}}$ consists of only Clifford gates.
  \item $C_T$ consists of only disjoint single-qubit $T$ gates.
  \item No gate from $C_{\text{after}}$ can be moved into $C_T$ or $C_{\text{before}}$ without violating the guarantees above.
\end{enumerate}
In a nutshell, sifting iterates over every gate in $C$ (in any valid time-ordering of the gates), and greedily adds each gate to the earliest of $(C_{\text{before}}, C_T, C_{\text{after}})$ possible without violating the partial time-ordering of gates~\footnote{As a minor improvement, we can allow the sifting algorithm to commute Clifford gates that are diagonal on a qubit with a $T$ gate on that qubit.}.
Recursively sifting the trailing circuits $C_{\text{after}}$ for $T$ gates yields a $T$-depth-optimal schedule for the gates in $C$.
Every $T$-layer can then be applied with one $T$-gate injection, while the Clifford layers in between can be resynthesized and injected according to the prescription in \cref{sec:Clifford}.

\clearpage
\section*{Disclaimer}

This paper was prepared for informational purposes with contributions from the Global Technology Applied Research center of JPMorgan Chase \& Co. This paper is not a product of the Research Department of JPMorgan Chase \& Co. or its affiliates. Neither JPMorgan Chase \& Co. nor any of its affiliates makes any explicit or implied representation or warranty and none of them accept any liability in connection with this paper, including, without limitation, with respect to the completeness, accuracy, or reliability of the information contained herein and the potential legal, compliance, tax, or accounting effects thereof. This document is not intended as investment research or investment advice, or as a recommendation, offer, or solicitation for the purchase or sale of any security, financial instrument, financial product or service, or to be used in any way for evaluating the merits of participating in any transaction.

\end{document}